\documentclass[aps,twocolumn,superscriptaddress,prr]{revtex4-1}

\usepackage{amssymb,graphicx,color,amsmath,xfrac,bm,stmaryrd,trimclip} 

\usepackage[utf8]{inputenc}

\definecolor{lightblue}{rgb}{0.17,0.39,1}
\usepackage[bookmarks, colorlinks=true, breaklinks]{hyperref}
\hypersetup{linkcolor=lightblue, citecolor=lightblue, filecolor=black, urlcolor=lightblue}

\newcommand{\LSCO}{La$_{2-x}$Sr$_x$CuO$_4$}

\begin{document}


\title{Universal Planckian dissipation in the strange metal state of the cuprates}


\author{A.~Shekhter }
\email[email: ]{arkadyshekhter@gmail.com}
\affiliation{Los Alamos National Laboratory, Los Alamos, NM 87545, USA}

\author{ B.~J.~Ramshaw }
\affiliation{Laboratory of Atomic and Solid State Physics, Cornell University, Ithaca, NY, USA}

\author{ M.~K.~Chan }
\affiliation{Los Alamos National Laboratory, Los Alamos, NM 87545, USA}

\author{N.~Harrison}
\affiliation{Los Alamos National Laboratory, Los Alamos, NM 87545, USA}

\begin{abstract}
A major puzzle in high-$T_{\rm c}$ superconductivity is the origin of the ``Planckian'' relaxation rate $1/\tau$ underlying the linear-in-temperature resistivity in the strange-metal state, which persists up to very high temperatures. Implicit in theoretical discussions is the assumption that $1/\tau$ must be universal. Experimentally, it is unclear, however, how such universality can be reconciled with the observed strong doping dependence of the resistivity over a wide doping range.  We show, through an analysis of a large body of optical conductivity and electrical resistivity data, that a universal $1/\tau$ requires only that the square optical plasma frequency $\omega_{\rm opt}^2(p)$ scales linearly with $p$ across the entire doping range, as is observed experimentally. We further argue that this can be understood via a Gutzwiller factor in doped Mott insulators of the form proposed by Anderson [\emph{Science} \textbf{235}, 1196 (1987)]. 
\end{abstract}


\maketitle



The discovery of high-temperature superconductivity in the cuprates~\cite{bednortz1986} was significant not only because of their unusually high superconducting transition temperatures, \( T_{\rm c} \), but also because their normal states represented the first clear example of a class of unconventional metals termed `strange metals'~\cite{anderson1988}, to distinguish them from conventional metals~\cite{landau1957}. Of particular interest is the electrical resistivity of strange metals~\cite{anderson1992,zaanen2004,keimer2015}, which unambiguously defies Fermi-liquid phenomenology~\cite{PinesNozieres1999}. Since the discovery of the cuprates, similar strange-metal behavior has been observed in a wide range of materials~\cite{bruin2013,hartnoll2022,phillips2022,cao2020}, most of which also exhibit unconventional superconductivity. Today, the research of strange metals is closely tied to the search for unconventional superconductors.

The strange metal state in the cuprates~\cite{anderson1988, Varma1989, nagaosa1992, anderson1992, phillips2022} is characterized by a linear-in-temperature (\( T \)-linear) resistivity~\cite{martin1990, ando2004, keimer2015}, expressed as \( \rho = A\,T \), that persists up to temperatures of \( T \sim 1000 \)~K---limited only by chemical stability of these compounds. This form of resistivity is `scale-invariant' because the only energy scale governing electrical transport is the externally imposed temperature, \( k_{\rm B} T \), rather than an intrinsic energy scale~\cite{wilson1974,lohneysen2007}. This stands in stark contrast to conventional metals, whose transport is governed by the Fermi energy—an intrinsic property of the electronic system---together with temperature~\cite{landau1957,PinesNozieres1999}.

While the microscopic mechanism of strange metal behavior remains heavily debated~\cite{Varma1989, anderson1992, emery1995i, keimer2015, zaanen2019, varma2020, hartnoll2022, phillips2022}, the \( T \)-linear resistivity implies the equivalence of two energy scales: the relaxation rate, \( \hbar/\tau \), and the thermal energy, \( k_{\rm B} T \):
\begin{equation}\label{planckianequation}
\hbar/\tau = \alpha\,k_{\rm B}T.
\end{equation}
This equivalency, where the factor $\alpha$ is of order unity, has been termed `Planckian' due to a purported naturalness of the connection between these two energy scales within several theoretical frameworks~\cite{Varma1989, zaanen2004, davison2014, zaanen2019,anderson1992, aji2007, phillips2022, hartnoll2022,patel2023,chang2025}. Implicit to these theoretical discussions is a doping {\it independence} of $\alpha$. In fact, values of $\alpha$ close to unity have been reported on the basis of low-temperature resistivity behavior at select dopings in several cuprate families~\cite{bruin2013,legros2019} (see Methods~\ref{prior}), including $p=$~0.26 in \LSCO~(LSCO). Because the slope \( A(p) \) of the \( T \)-linear electrical resistivity in the strange metal state of the cuprates is strongly doping-dependent~\cite{takagi1992, ito1993, watanabe1997, ando2004, barisic2013i,giraldogallo2018}, it is not at all clear
how the purported universal value of $\alpha$ can be reconciled with the observed resistivity behavior throughout the phase diagram. 

Existing approaches to analyzing the electrical resistivity in the context of Planckian dissipation use a Drude model~\cite{cooper2009,bruin2013,legros2019,proust2019}, in which the electrical conductivity is expressed in terms of the ratio of the carrier density \(n\) to an effective mass \(m^\ast\)~\cite{ashcroft1976}. Such an approximation is warranted for isotropic systems, especially those with a parabolic band, for which Galilean-invariance is either an intrinsic or emergent property (see Methods~\ref{lack}). In these cases, both the electronic specific heat coefficient \(\gamma\) (and the associated effective mass) and the Hall coefficient \(R_{\rm H}\) are expected to be only weakly temperature dependent~\cite{landau1957,PinesNozieres1999}.

In the cuprates, however, \(\gamma(p,T)\)~\cite{loram2001,tallon2022} and \(R_{\rm H}(p,T)\)~\cite{ando2004a,ono2007} exhibit strong, and in some cases non-monotonic, dependences on both temperature and doping. At the very least, these behaviors reflect a pronounced four-fold anisotropy in the electronic structure, arising in part from proximity to a van Hove singularity~\cite{horio2018} and from pseudogap effects~\cite{timusk1999}. Because of this anisotropy,  the electrical transport in the cuprates must be understood in terms of Boltzmann transport~\cite{grissonnanche2021,ataei2022}, rather than through the Drude approximation. Surprisingly, an approach based on Boltzmann transport has not yet been applied to the question of universal Planckian dissipation.

Boltzmann transport supersedes the Drude approximation because it correctly incorporates anisotropies and singularities in the electronic dispersion~\cite{ashcroft1976}. The experimentally relevant quantity for zero-field transport is the Fermi-surface velocity average entering the squared optical plasma frequency, \(\omega_{\rm opt}^2\) (see Methods~\ref{lack} and~\ref{calculating}). This enables the universality of the Planckian relaxation rate across the cuprate phase diagram to be addressed in two important ways. First, because the conductivity is weighted toward regions of the Fermi surface with large Fermi velocity, the in-plane conductivity of the cuprates is weighted primarily toward the nodal states~\cite{grissonnanche2021,ataei2022}, which remain comparatively robust across the phase diagram~\cite{yoshida2006}. This removes the need, inherent in the Drude approximation, to consider \(n\) and \(m^\ast\), both of which have been argued to exhibit strong and non-monotonic dependences on doping and temperature~\cite{loram2001,momono2002,ando2004,badoux2016,laliberte2016,michon2019,putzke2021,girod2021,shekhter2022,legros2022}. Second, because \(\omega_{\rm opt}^2\) is determined independently using optical conductivity~\cite{padilla2005,legros2022,michon2021}, it enables a direct comparison between this quantity and the inverse slope of the \(T\)-linear resistivity.

Here, we show that a comparison between electrical transport data and the optically determined \(\omega_{\rm opt}^2\)~\cite{padilla2005,legros2022,michon2021} enables
a quantitative analysis of the Planckian relaxation rate in the cuprates to be made across the phase diagram. A striking observation is that $\omega_{\rm opt}^2(p)$ varies linearly with doping (i.e., is $p$-linear) over the entire doping range, suggesting a true universality of Planckian dissipation. The observed $p$-linear doping dependence of \(\omega_{\rm opt}^2(p)\) is such a conspicuously simple form that it begs its own explanation. We address this question in the second half of the paper by showing that the observed behavior arises naturally within a Mott picture~\cite{vollhardt1984,vollhardt1987}, in which the electronic dispersion is renormalized by a Gutzwiller factor \(q \approx p\)~\cite{anderson1987}. This theoretical framework can be tested not only through  \(\omega_{\rm opt}^2(p)\), but also through a comparison with entropy measurements.


\section*{Results}

\subsection*{Evidence for universality of the Planckian relaxation rate} 

\begin{figure}[t!!!!!!!!!!!] 
\begin{center}
\includegraphics[width=0.9\linewidth]{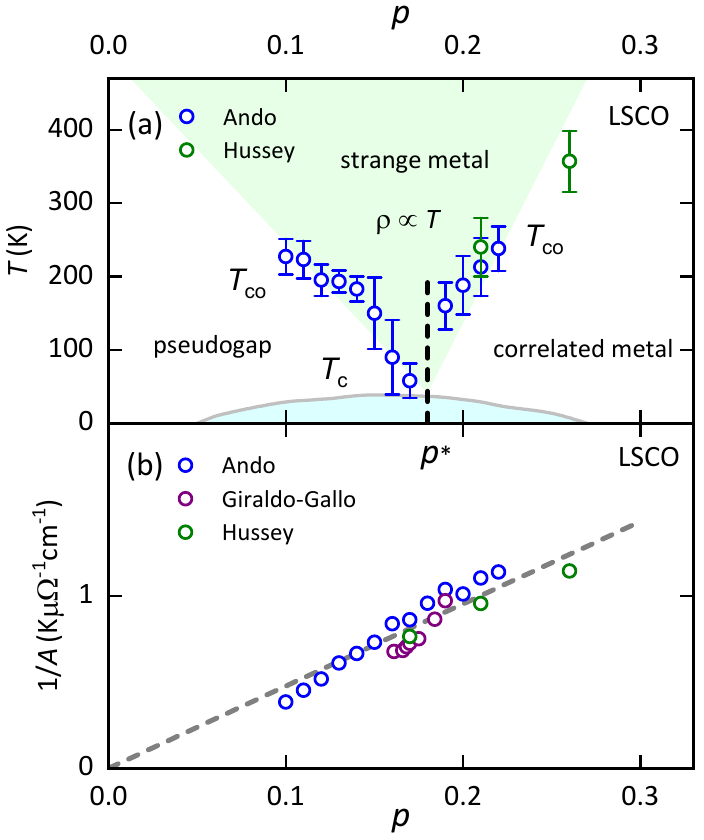}
\caption{
(a) Strange metal phase (shaded green) where $\rho\approx\rho_0+AT$~\cite{ando2004}, within which $\rho_0$ refers $\rho$ extrapolated to $T=0$.  $T_{\rm co}$ (circles) is obtained by fitting data of Ando {\it et al.} (see Methods~\ref{linear}), or from the saturation of the temperature derivative $\partial\rho(T)/\partial T$ at high temperatures, as shown by Hussey {\it et al.}~\cite{hussey2011}.  For $p>$~0.2, observation of saturation requires measurements at least to $\approx$~400~K.  
 (b) Corresponding high temperature saturated $1/A$ values (circles) determined
by Hussey {\it et al.}~\cite{hussey2011,legros2019}, and Giraldo-Gallo {\it et al.}~\cite{giraldogallo2018}, and by performing linear regression of data from Ando~{\it et al.}~\cite{ando2004} (see Methods~\ref{linear}), Fig.~\ref{andofits}). In all cases, we have only included $A(p)$ values explicitly indicated in the raw $\rho(T)$ or $\partial\rho(T)/\partial T$ data. 
Note that for $p\lesssim$~0.1, $\rho$ vs. $T$ is no longer linear~\cite{ando2004,ono2007,takagi1992}, possibly as the result of antiferromagnetic and spin glass phases at these dopings. The dashed line in a guide to the eye.}
\label{andoplot}
\end{center}
\end{figure}

Figure~\ref{andoplot}(a) shows the phase diagram of LSCO indicating the strange metal state above a crossover temperature $T_{\rm co}(p)$, within which region the resistivity is $\rho\approx\rho_0+A(p)T$, where $\rho_0$ refers to resistivity extrapolated to $T=0$. While the strange metal state extends down to low temperatures near the `critical doping' \( p^\ast \approx 0.18 \)~\cite{mackenzie1996, ando2004, cooper2009, keimer2015}, as one moves away from \( p^\ast \), the high-temperature \( T \)-linear behavior is cut off below \( T_{\rm co}(p) \)~\cite{hussey2011} [Fig.~\ref{andoplot}(a)]. Below this temperature, the resistivity is typically superlinear and no longer scale-invariant because it depends on energy scales other than temperature (see Methods~\ref{prior}). Here, we focus on the strange metal state exhibiting scale-invariant (i.e. $T$-linear) resistivity, which is confined to the funnel-shaped region of the phase diagram above \( T_{\mathrm{co}}(p) \). Figure~\ref{andoplot}(b) shows the inverse slope, $1/A(p)$, of the $T$-linear resistivity obtained by linear regression of $\rho(T)$ at high temperatures in LSCO over a broad range of dopings~\cite{ando2004,giraldogallo2018,hussey2011} (see Methods~\ref{linear}). As shown in Fig.~\ref{andoplot}(b), \( 1/A(p) \) within the strange metal state increases monotonically with \( p \) over a wide doping range (\( 0.1 \leq p \leq 0.26 \)), suggesting that $A(p)\propto1/p$ (see dashed line).

The electrical transport can be analyzed within the framework of the Einstein relation for conductivity, $\sigma = e^2 \kappa D_{\rm e}$, which has been extensively applied to studies of two-dimensional low-carrier density electron liquids~\cite{finkelstein1990}. For parabolic bands typical for such low-carrier density systems, the conductivity $\sigma = 1/\rho$ is expressed as the product of the electronic compressibility $\kappa = \partial n / \partial \mu$ and the diffusion coefficient $D_{\rm e} = v_{\rm F}^2 \tau_{\rm e}/ 2$ at the Fermi surface, where $v_{\rm F}=\hbar^{-1}(\partial\epsilon_k/\partial k)_{k=k_{\rm F}}$ is the Fermi velocity.  When Fermi-liquid interaction effects are included, the electronic compressibility is renormalized by a dimensionless Landau interaction factor: $\kappa = (1 + F_0^{\rm s})^{-1} \nu$~\cite{landau1957}, where $\nu = m^\ast / \pi \hbar^2$ is the electronic density of states (assuming a two-dimensional system). Here, $F_0^{\rm s}$ represents the zeroth harmonic Landau interaction parameter in the charge channel. 

The effective mass near the Fermi surface is defined as $m^\ast=\hbar\,k_{\rm F}/v_{\rm F}$, where $\hbar\,k_{\rm F}$ is the Fermi momentum, and is directly determined by the electronic specific heat. 
Meanwhile, the relaxation time entering the diffusion coefficient is an effective quantity that incorporates Fermi-liquid interaction factors, such that $\tau_{\rm e} = \tau (1 + F_0^{\rm s})$~\cite{finkelstein1983,finkelstein1990}, because it characterizes collective dynamics of particle-hole diffusion rather than the single-particle relaxation rate. Note here also that $\tau_{\rm e}$ does not represent the characteristic time for quasiparticle decay~\cite{stricker2014}. It is well known that in the zero-frequency limit of the conductivity, the Fermi-liquid interaction factors cancel. Hence, $(1 + F_0^{\rm s})$ renormalizes both the compressibility and the effective scattering rate. 

The van Hove singularity (at \( p \approx 0.20 \) in LSCO~\cite{horio2018}) represents a particularly strong deviation from band parabolicity in the cuprates that cannot be neglected. Its effect is not confined to a single doping level, but extends across the entire doping range of interest. To take into consideration this effect, the Einstein relation must be recast in a form consistent with Boltzmann transport~\cite{ashcroft1976}: $\sigma = e^2 \langle \kappa(k) D_{\rm e}(k) \rangle$, where $\kappa(k)$ and $D_{\rm e}(k)$ denote the partial compressibility and diffusion coefficient at each momentum $k$ on the Fermi surface, respectively, and $\langle \cdots \rangle$ indicates an average over the Fermi surface. 
Note that because Planckian dissipation is a local phenomenon~\cite{zaanen2020}, the resulting Planckian relaxation rate is expected to be the same everywhere on the Fermi surface, consistent with experimental observations~\cite{grissonnanche2021}.

\subsection*{Optical conductivity evidence for universality} 

In optical conductivity measurements, the low-frequency part of the optical conductivity, referred to as the Drude peak, is analyzed using the Lorentzian  form~\cite{PinesNozieres1999, basov2005, armitage2009}
\begin{equation}\label{frequencydependent}
\sigma(\omega) = (\omega_{\rm opt}^2 / 4\pi) \times 1 / (\tau^{-1}_{\rm e} - i\omega).
\end{equation}
In the limit $\omega\to0$, this correctly reproduces the dc-limit conductivity: $\sigma=(\omega^2_{\rm opt}/4\pi)\tau_{\rm e}$. Hence,
\begin{align}\label{zerofrequencyoptics}
\sigma = \left( \frac{\omega_{\rm opt}^2}{4\pi} \right) \tau_{\rm e},
\end{align}
which must be compared to $\sigma = e^2 \langle \kappa(k) v_x^2(k) \rangle \tau_{\rm e}$ from the Einstein relation (see Methods~\ref{calculating}). Therefore, $\omega^2_{\rm opt}/4\pi$ is defined in terms of the charge compressibility, which includes the Fermi liquid interaction factor, $\omega^2_{\rm opt}/4\pi=e^2 \langle \kappa(k) v_x^2(k) \rangle$.

\begin{figure}[t!!!!!!!!!!!] 
\begin{center}
\includegraphics[width=0.9\linewidth]{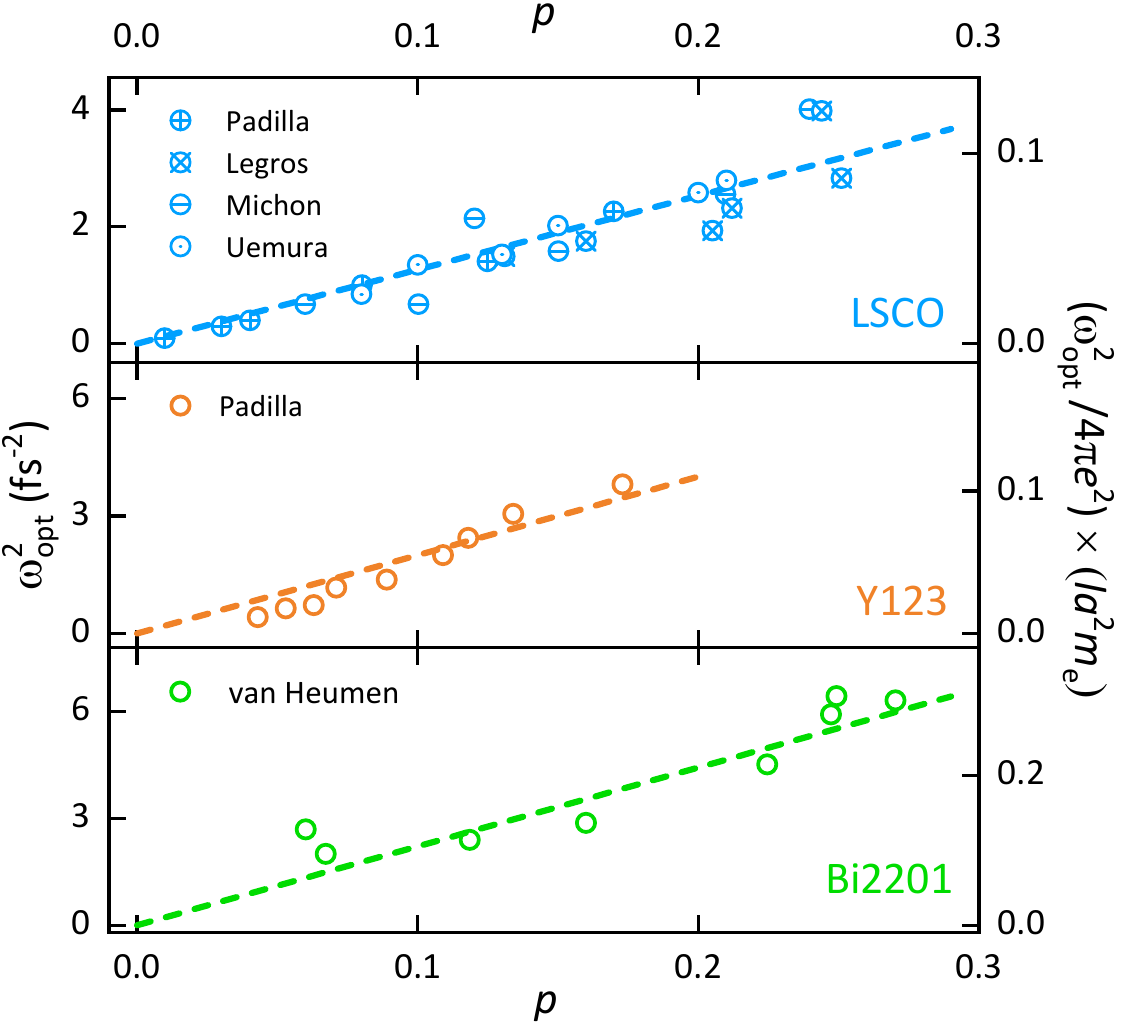}
\caption{$\omega_{\rm opt}^2(p)$ for LSCO from Padilla {\it et al.}~\cite{padilla2005}, Legros {\it et al.}~\cite{legros2022}, and Michon {\it et al.}~\cite{michon2021}; for Y123 from Padilla {\it et al.}~\cite{padilla2005} with doping $p$ determined by Liang {\it et al.}~\cite{liang2006}; and for Bi2201 from van Heumen {\it et al.}~\cite{vanheumen2022}. For LSCO, we also include estimates of $\omega_{\rm opt}^2$ inferred from penetration-depth measurements by Uemura {\it et al.}~\cite{uemura1989, uemura1991}, up to $p \!\sim\! 0.2$ where $p$-linearity is observed, which fall on the same line. Note that the optical studies above determine $\omega_{\rm opt}^2(p)$ from the spectral weight (area) of the Drude peak. However, separate values of $n$ and $m^\ast$ instead of $\omega_{\rm opt}^2(p)$ are sometimes reported under the assumption of a parabolic band dispersion~\cite{padilla2005,legros2022} (and neglecting $(1+F_0^{\rm s})$). In contrast, the present work focuses on $\omega_{\rm opt}^2(p)$, whose determination is independent of any parabolic band assumption. This requires us to reconstruct the measured $\omega_{\rm opt}^2(p)$ from the published $n$ and $m^\ast$ values using the specific form of the relation $\omega_{\rm opt}^2(p) = 4\pi e^2 n / m^\ast$ used in Refs.~\cite{padilla2005,legros2022}.
}
\label{drudeweight}
\end{center}
\end{figure}

While it is quite possible that the effective relaxation rate $1/\tau_{\rm e}$ is frequency dependent, in the cuprates there is as yet no clear experimental evidence for a significant departure from a Lorentzian lineshape for frequencies in the regime $\omega \lesssim 1/\tau_{\rm e}$~\cite{vanheumen2022}. The possible frequency dependence we are considering here pertains to changes in the character of Fermi liquid renormalizations of $\tau_{\rm e}$ rather than intrinsically frequency dependent quasiparticle lifetime effects~\cite{stricker2014}. At frequencies $\omega>1/\tau_{\rm e}$, meanwhile, the Drude response is strongly suppressed, and the optical conductivity becomes dominated by the mid-infrared optical response~\cite{basov2005}. 

Experimentally, the prefactor $\omega_{\rm opt}^2 / 4\pi$ is determined from the area under the Drude peak. Because the Drude peak is only clearly resolved in the cuprates as a contribution distinct from from the mid-infrared background only for $\omega \lesssim \tau_{\rm e}^{-1}$, $\omega_{\rm opt}^2 / 4\pi$ is obtained from fitting~\cite{padilla2005,basov2005}. Therefore, the value of $\omega_{\rm opt}^2 / 4\pi$ is entirely represented by the central part of the Drude peak, within which $1/\tau_{\rm e}$ continues to adequately characterize its width. This implies that the optical plasma frequency obtained in this way reflects the electronic compressibility, which includes the Fermi-liquid interaction factor. Using Eqs.~\ref{planckianequation} and~\ref{zerofrequencyoptics}, the slope $A(p)$ of the $T$-linear resistivity can then be expressed as
\begin{equation}\label{coefficient}
A(p) = \left( \frac{4\pi}{\omega_{\rm opt}^2(p)} \right) \left( \frac{k_{\rm B}}{\hbar} \right) \alpha(p)\,\times\,(1 + F_0^{\rm s})^{-1},
\end{equation}
where the doping dependence may arise from either $\alpha(p)$ or $\omega_{\rm opt}^2(p)$. For $\alpha$ to be doping independent (i.e. universal), the observed doping dependence of $A(p)$ (approximately $\propto1/p$ in Fig.~\ref{andoplot}) must originate {\it entirely} from $\omega_{\rm opt}^2(p) \propto p$.

A $p$-linear scaling  \( \omega_{\rm opt}^2(p)\propto p \) is indeed readily observed in optical conductivity measurements. Figure~\ref{drudeweight} shows the plasma frequency squared in LSCO~\cite{padilla2005}, YBa$_2$Cu$_3$O$_{6+x}$ (Y123)~\cite{padilla2005}, and Bi$_{2-x}$Pb$_x$Sr$_{2-y}$La$_y$CuO$_{6+\delta}$ (Bi2201)~\cite{vanheumen2022}. While the slopes differ somewhat—possibly reflecting variations in electronic structure—all three systems exhibit $\omega_{\rm opt}^2(p) \propto p$ behavior up to $p\approx$~0.3.  

Given Eq.~\ref{coefficient}, optical conductivity and transport measurements together provide model-independent experimental evidence for the doping independence of $\alpha$ (i.e., the universality of the Planckian relaxation rate). This directly addresses the main question raised in the introduction. The remainder of this manuscript is devoted to a discussion of a possible underlying mechanism responsible for the relation $\omega_{\rm opt}^2(p) \propto p$, which we start by turning to thermodynamic measurements.

\subsection*{Entropic evidence for $\omega_{\rm opt}^2\propto p$ at higher dopings} 

\begin{figure}[t!!!!!!!!!!!]
\begin{center}
\includegraphics[width=0.9\linewidth]{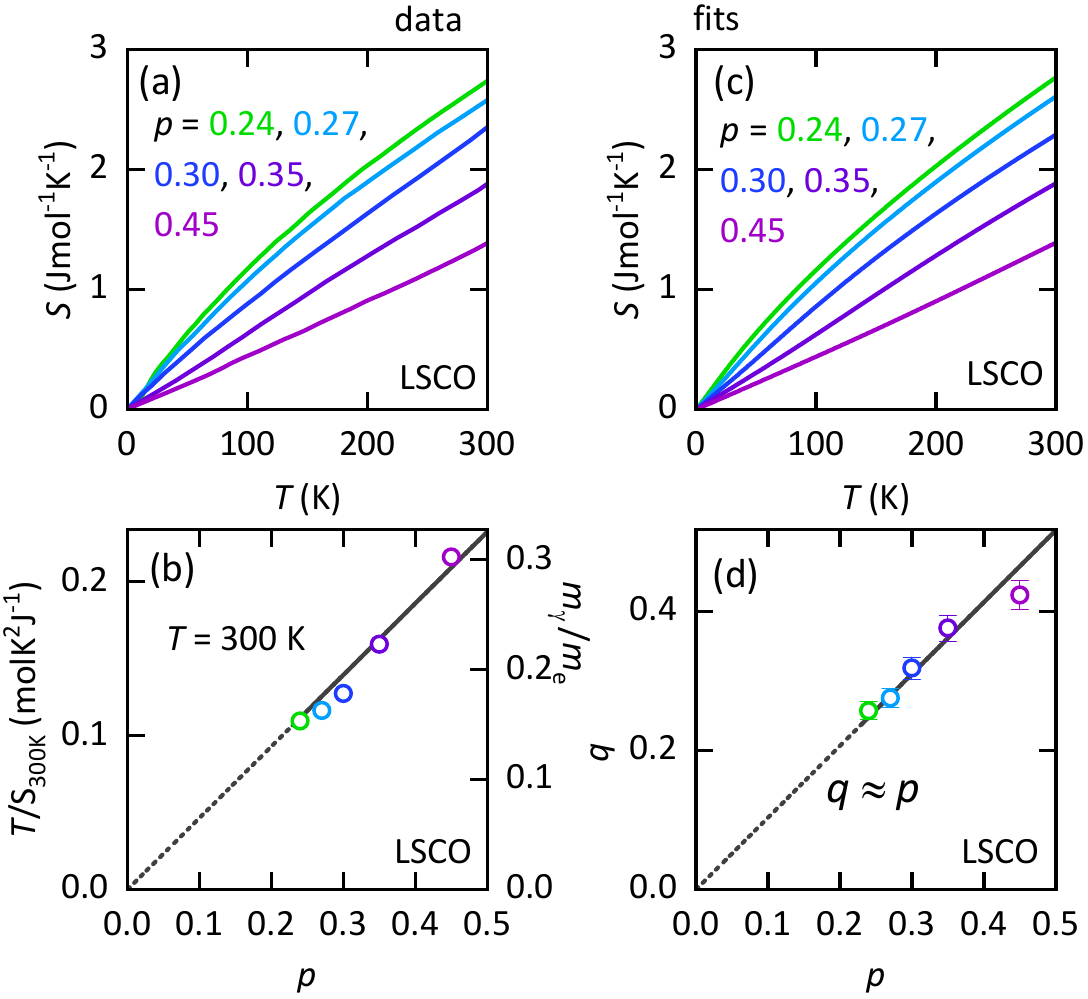}
\caption{ 
(a) Measured $S$ vs. $T$ of LSCO at several $p$~\cite{tallon2022}. (b) $T/S_{\rm 300K}$ vs. $p$ (lines rescaled from (d). (c) Calculated $S$ vs. $T$  using $q$ values (circles) from (c) (see Methods~\ref{tight} and~\ref{determining}). (d) Fitted $q$ vs. $p$, where the solid line corresponds to $q=(1.03\pm0.02)\times p$. The dotted line is an extrapolation to $p=0$. 
}
\label{tallonplot}
\end{center}
 \end{figure}

Thermodynamic measurements (Fig.~\ref{tallonplot})~\cite{loram2001}, which predate the optical studies, provide independent experimental scrutiny of the linear scaling $\omega_{\rm opt}^2(p) \propto p$ obtained from optical conductivity measurements, particularly at higher dopings (\( p \geq 0.23 \)) where there are no apparent pseudogap effects. With increasing doping~\cite{loram2001, tallon2022}, the high-temperature entropy \( S(T) \) is approximately linear in temperature [Fig.~\ref{tallonplot}(a)], allowing the specific heat in mass units $m_\gamma$ to be approximately estimated directly from the electronic specific heat coefficient  \( \gamma  \approx S(T)/T \) via 
\begin{equation}\label{rourke}
\gamma = m_\gamma \left( \frac{\pi k_{\rm B}^2 N_{\rm A} a^2}{3 \hbar^2} \right)
\approx \frac{m_\gamma}{m_{\rm e}} \times (1.4~\text{J mol}^{-1} \text{K}^{-2}).
\end{equation}
This form assumes a quasi-two-dimensional electronic structure~\cite{rourke2010}, where $m_{\rm e}$ is the free-electron mass. Figure~\ref{tallonplot}(b) shows \( \gamma \approx S(T)/T \) at \( T = 300~\text{K} \), and that the inferred $m_\gamma(p)$ scales as \( 1/p \) for dopings \( 0.23 \leq p \leq 0.45 \), spanning a factor of two in $p$.

It is noteworthy that the observed $1/p$ behavior of $m_\gamma$ extends that of $\omega_{\rm opt}^2(p)$ in Fig.~\ref{drudeweight} to higher dopings. In fact, the $p$-inverse scaling $m_\gamma(p) \propto1/p$  implies $\omega_{\rm opt}^2(p) \propto p$ in the $p>$~0.23 regime. This relation can be quantitatively demonstrated by adopting a phenomenological model of a `Mott band' that reproduces the experimentally observed $m_\gamma(p) \propto1/p$ behavior in Fig.~\ref{tallonplot}(b). In this model, the Mott band corresponds to an empirically-validated~\cite{horio2018} tight-binding dispersion (see Methods~\ref{tight}), where the nearest neighbor hopping predicted by density functional theory is rescaled by a Gutzwiller factor $q$~\cite{brinkman1970}. We determine $q$ at each doping by comparing the calculated electronic entropy with that measured~\cite{loram2001,tallon2022} over a broad temperature range (see Methods~\ref{determining}). At high doping levels, we find that 
\begin{equation}\label{massrenormalization}
    q\approx p\,,
\end{equation}
which is consistent with the near-linear dependence of the high-temperature entropy on $q$ [see Fig.~\ref{tallonplot}(b)].  Figure~\ref{tallonplot}(c) shows the entropy calculated using such a model, while Fig.~\ref{tallonplot}(d) shows $q$ determined at each doping together with a straight line through the points depicting Equation~\ref{massrenormalization}.

\begin{figure}[t!!!!!!]
\begin{center}
\includegraphics[width=0.9\linewidth]{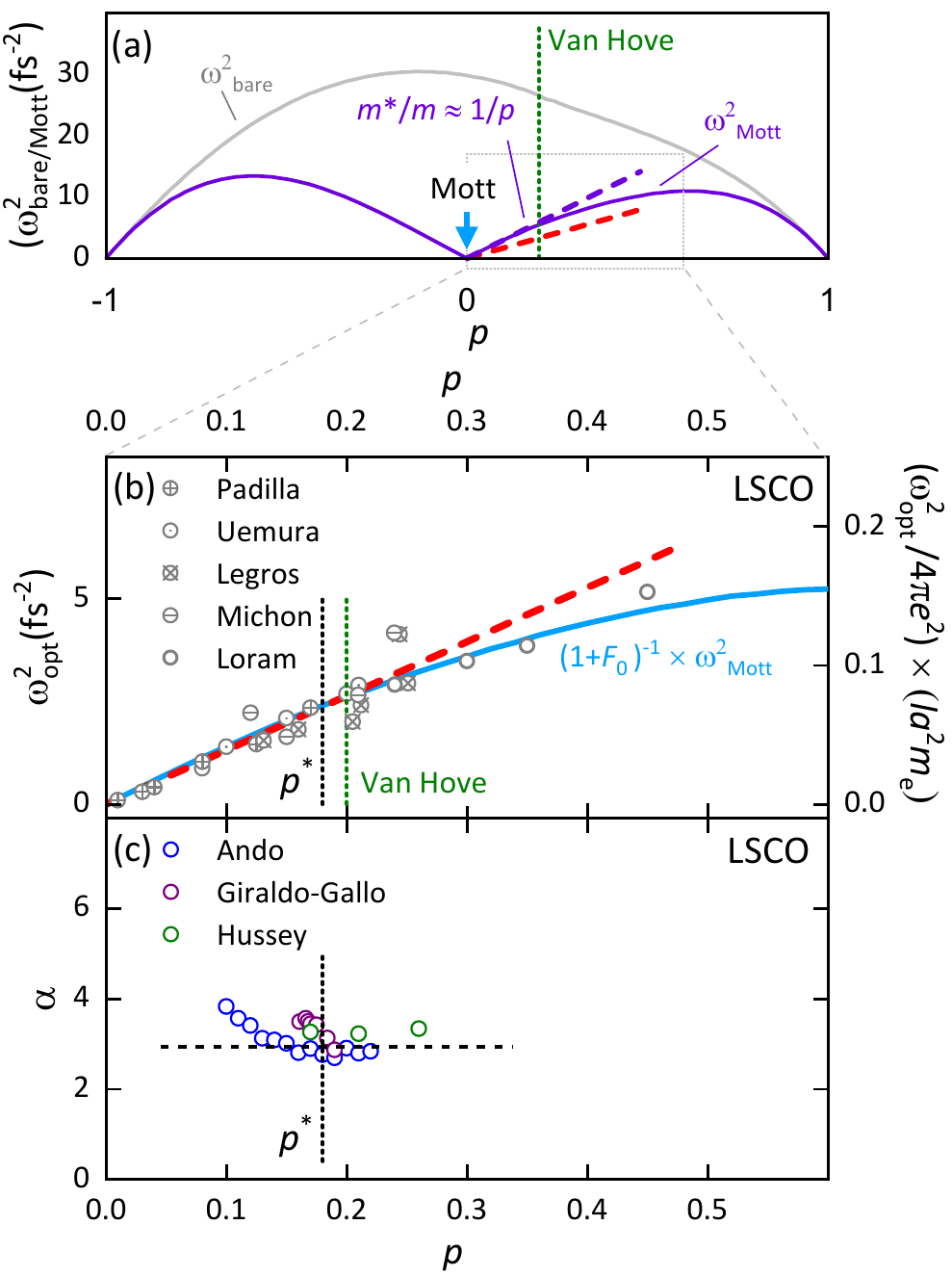}
\caption{
(a) Calculated $\omega_{\rm bare}^2$ (grey curve) according to the tight-binding parameters of Horio~{\it et al.}~\cite{horio2018} (see Methods~\ref{calculating} and~\ref{tight}), and $\omega_{\rm Mott}^2(p)$ (purple curve) using $q$ given by Eq.~\ref{massrenormalization}. A similar doping-dependent Mott band was recently discussed in Refs.~\cite{phillips2020,zaanen2020}. The purple dashed line is its $p\rightarrow0$ asymptote. The red dashed represents the data from panel (b). (b) Measured $\omega_{\rm opt}^2$  in LSCO from the measured optical conductivity of Padilla {\it et al.}~\cite{padilla2005}, time-resolved THz spectroscopy of Legros {\it et al.}~\cite{legros2022}, and penetration depth of Uemura {\it et al.}~\cite{uemura1989,uemura1991}. We also show $\omega_{\rm opt}^2 = \omega_{\rm Mott}^2\times(1 + F_0^{\rm s})^{-1}$ using $q\approx p$ (blue curve) and using the experimentally determined values of $q$ from Fig.~\ref{tallonplot}(d). 
(c) Values of $\alpha$ determined using the data in Fig.~\ref{andoplot}(b), Eq.~\ref{coefficient}, and the red dashed line corresponding to $(\omega^2_{\rm opt}/4\pi e^2)\times(la^2m_{\rm e})=0.39\,p$. 
}
\label{novermsimple}
\end{center}
 \end{figure}

Equation~\ref{massrenormalization}  enables a renormalized squared plasma frequency of the Mott band $\omega^2_{\rm Mott}(p)$ to be estimated at all dopings from the tight-binding approximation of a Mott band (see Methods~\ref{calculating}), as indicated in Fig.~\ref{novermsimple}(a) (purple line). The purple dashed line shows the asymptotic behavior of $\omega_{\rm Mott}^2(p)$ as $p \rightarrow 0$, while the red dashed line represents the estimated $p$-linear behavior of $\omega_{\rm opt}^2(p)$ in LSCO, as extracted from optical conductivity data in Fig.~\ref{drudeweight} [replotted in Fig.~\ref{novermsimple}(b)]. 

It is interesting that there is a clear difference in slope between the observed $\omega^2_{\rm opt}$ (red dashed line) and that $\omega^2_{\rm Mott}$ (purple dashed line) expected from the specific heat of a phenomenological Mott band. Such a difference in slope is in fact anticipated if one recalls that the estimate $\omega_{\rm Mott}^2(p)$, derived from the entropy-informed tight-binding model, uses the density of states $\nu$ (constrained by entropy measurements), whereas $\omega^2_{\rm opt}$ depends on $\kappa = \partial n / \partial \mu$. As discussed earlier, these differ by the Landau interaction factor; i.e., $\kappa=(1+F_0^{\rm s})^{-1}\,\nu$~\cite{landau1957,leggett1968,leggett1975}. Since the entropy-constrained $\omega_{\rm Mott}^2(p)$ does not include the $(1+F_0^{\rm s})^{-1}$ factor, we expect the ratio \( \omega_{\rm opt}^2/\omega_{\rm Mott}^2 = (1+F_0^{\rm s})^{-1}  \), suggesting a modest Fermi liquid interaction factor  $(1+F_0^{\rm s})^{-1} \approx 0.6$ across the entire doping range. In Fig.~\ref{novermsimple}(b), we include estimates of $\omega_{\rm opt}^2$ obtained from the individual experimentally determined values of $q$ shown in Fig.~\ref{tallonplot}(d) at each doping level.

It is important to note that while the estimated \( \omega_{\rm opt}^2(p) \) remains $p$-linear up to \( p \sim 0.3 \) [Figs.~\ref{novermsimple}(a) \& (b)], it begins to bend downward on approaching the band extremity at $p=1$ [Fig.~\ref{novermsimple}(a)], driven by the decrease in electrons density in this limit. This behavior was recently highlighted in an analysis of the superfluid density in doped Mott insulators~\cite{phillips2020,zaanen2020}. The slight asymmetry in Fig.~\ref{novermsimple}(a) compared to Refs.~\cite{phillips2020,zaanen2020} is the result of \textcolor{black}{next-nearest neighbor hopping} parameters specific to LSCO~\cite{horio2018}. 

The existing data for \(1/A(p)\propto p\) in Fig.~\ref{andoplot}(a) extends only up to \( p = 0.26 \), where \(\omega_{\rm opt}^2(p)\) is still a linear function of $p$. We therefore use Eq.~\ref{coefficient} to obtain \( \alpha \approx 3 \)~\cite{chang2025}, corresponding to the median value indicated by the horizontal dashed line in Fig.~\ref{novermsimple}(c). Taken together, the optical conductivity, penetration depth, entropy measurements, and \(T\)-linear resistivity in the strange metal state establish the doping-independence of $\alpha$---i.e., universality of the Planckian relaxation rate $1/\tau$ over the entire doping range analyzed, 0.1~$<p<$~0.3, and quite possibly outside it.

\subsection*{Mott physics as a possible mechanism for $\omega_{\rm opt}^2\propto p$} 

It is interesting to note that a similar Gutzwiller factor effects were previously identified in $^3$He~\cite{wheatley1975,leggett1975}. 
While it was initially suggested that the observed pressure-dependence of the Landau parameters in $^3$He were due to proximity to a Stoner instability~\cite{doniach1966}, it was subsequently shown that such an instability could not simultaneously account for the observed magnitude of the compressibility and magnetic susceptibility~\cite{armitage1975helium,vollhardt1984,vollhardt1987}. Specifically, it was  shown that Fermi liquid relations require the observed thermodynamic properties to be determined by an enhancement of the quasiparticle effective mass that occurs {\it independently} of Fermi liquid interactions.

The transformative idea that enabled a quantitative understanding of the pressure-dependent Fermi liquid parameters in $^3$He is that the enhancement of the quasiparticle effective mass is caused by the suppression of the mobility of $^3$He atoms by hard-core repulsive interactions. This is referred to as an `almost-localized Fermi liquid,' in analogy with a Mott insulator doped away from half-filling by a small amount $p$~\cite{armitage1975helium,vollhardt1984,vollhardt1987}.

To account for the strong hard-core repulsion between atoms in $^3$He, Anderson and Vollhardt~\cite{armitage1975helium,vollhardt1984,vollhardt1987} introduced the `lattice-gas model' with an on-site repulsion energy parameter $U$. A key feature of this model is that the number of lattice sites differs from the number of atoms~\cite{andreev1969}. Consequently, the liquid is described by a new thermodynamic parameter $p$ (denoted $\delta$ in Refs.~\cite{vollhardt1984,vollhardt1987}), akin to an order parameter without symmetry breaking~\cite{grilly1971}, which exists alongside pressure and temperature. This parameter defines a small difference between the volume per site of the lattice, $v_{\rm s}$, and the volume per atom, $v_{\rm a}$, expressed as $p = 1 - v_{\rm s} / v_{\rm a}$. Thus, it acts as an effective doping. Similar to the Landau description of phase transitions~\cite{landau1980}, $p$ itself is determined by a minimum in the free energy. While the Mott-insulating state occurs in this model at $ p=0 $, it is preempted in $^3$He by solidification at $p\approx 0.05$, corresponding to a pressure of 34.36 bar~\cite{grilly1971,vollhardt1987}.

In the limit of strong hard-core repulsion ($U\rightarrow\infty$), the effective mass enhancement in $^3$He is inversely proportional to the doping: $m/m^\ast=q \approx 1/2p$~\cite{vollhardt1984,vollhardt1987}. Such a comprehensive understanding of $^3$He~\cite{armitage1975helium,vollhardt1984,vollhardt1987} was predicated on the fact that the effective mass enhancement in \(^3\)He is driven by hard-core repulsion at short distances, which is the hallmark of Mott physics. In the cuprates, a discussion of the Mott-enhancement ratio \( q=m/m^\ast \) requires consideration of the fact that $m$ refers to a band mass~\cite{mattheiss1987,singh1994,pavarini2001} that is itself doping-dependent (see Methods~\ref{tight}). This is accounted for by a tight-binding approximation~\cite{horio2018} to the central Mott band. The empirically-determined Equation~\ref{massrenormalization} might therefore be taken as evidence of underlying Mott physics in the cuprates. The approximately twofold larger enhancement factor in LSCO, as expressed by Eq.~\ref{massrenormalization}, was in fact anticipated by Anderson, who extended his analysis of \(^3\)He to the cuprates using a modified argument~\cite{anderson1987,zou1988}.

In a $^3$He-inspired Mott scenario, the relation \( q \approx p \) follows from strong on-site repulsion $U$, which suppresses double occupancy and produces a narrow Mott band separated from upper and lower Hubbard bands by a Mott gap~\cite{vollhardt1984,vollhardt1987}. Because this form of the dispersion featuring a Mott gap is not continuously connected to the broad non-interacting band, the self-energy in the effective description of the central Mott band in terms of the Gutzwiller factor $q$ does not connect continuously to the non-interacting band when $U$ is decreased.

Phenomenological relations such as \( \tau^\ast = \tau (m/m^\ast) \)~\cite{PinesNozieres1999,basov2005,armitage2009,basov2011} in fact depend intrinsically on the analytic continuity of the interacting system to the non-interacting Fermi gas~\cite{landau1957,luttinger1960,leggett1968,leggett1975}. Such continuity is broken by the Mott physics discussed here and therefore such relations do not apply to the discussion here.  On the other hand, interaction factors that do not require adiabatic continuity to the non-interacting ($U=0$) limit, including \( \kappa = (1 + F_0^{\rm s})^{-1} \nu \) and \( \tau_{\rm e} = \tau (1 + F_0^{\rm s}) \), remain valid~\cite{finkelstein1990,vollhardt1984,vollhardt1987}. Because the Mott state is separated from the non-interacting ($U=0$) gas by a singularity, relations based on electron–phonon self-energy analogies~\cite{basov2011}, which themselves rely on analytic continuity of the phonon self-energies to the non-interacting system, do not apply to the effective Mott band at low energies.



\subsection*{$\omega_{\rm opt}^2\propto p$  across a Van Hove singularity }

\begin{figure}[t!!!]
\begin{centering}
\includegraphics[width=0.9\linewidth]{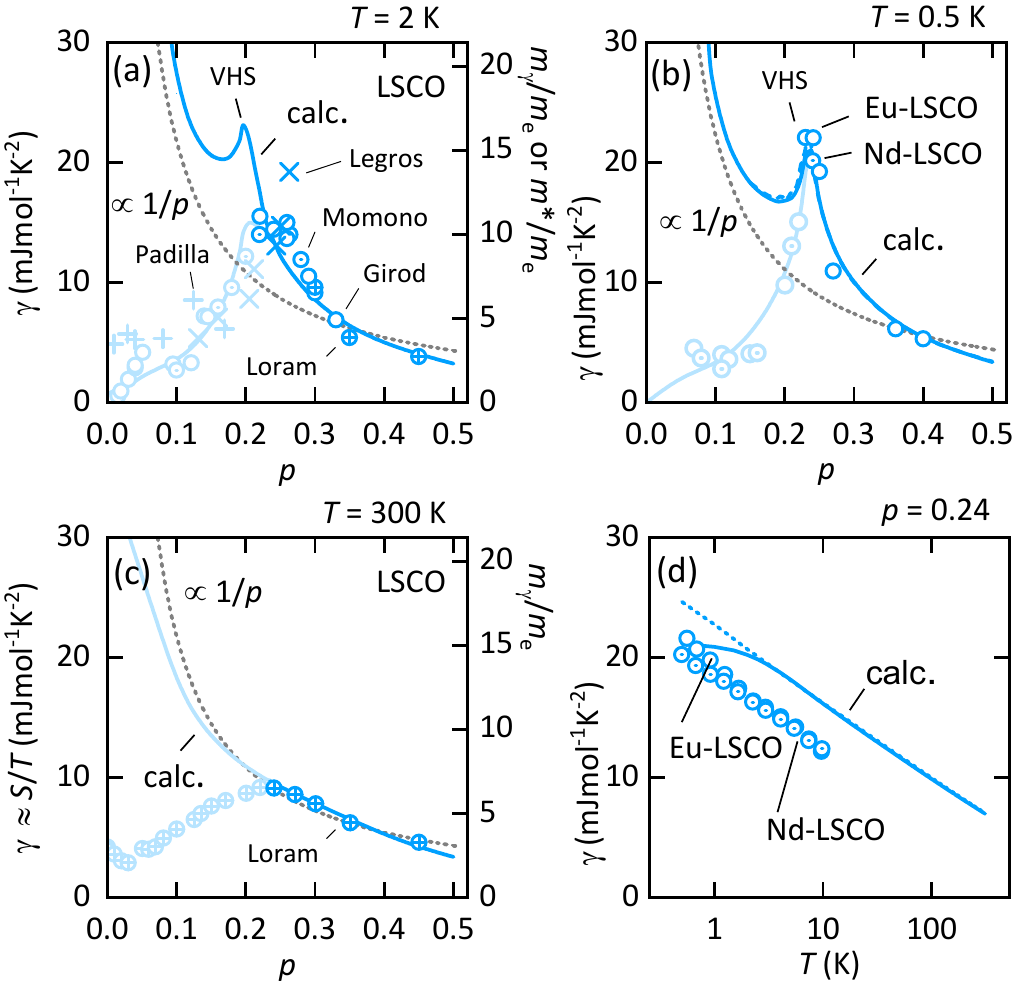}
\caption{ 
(a) $\gamma$ vs. $p$ from Refs.~\cite{loram2001,momono2002,girod2021} and equivalent $m_\gamma$ (right-hand axis), and also $m^\ast$ vs. $p$ from Ref.~\cite{padilla2005} ($+$~symbols) and Ref.~\cite{legros2022} ($\times$~symbols).
The solid line shows calculations of $\gamma=\partial S/\partial T$ (lines) including the Van Hove singularity (VHS) at $T=$~2~K  (the lowest $T$ accessed for $\gamma$ measurements~\cite{girod2021}) using $q=$~1.03$\times p$ and $t_z/t$ of 0.07~\cite{horio2018} (see Methods~\ref{tight} and~\ref{effect}). Points in the pseudogap regime ($p\lesssim$~0.23) are shaded lighter for clarity. The light blue line is a guide to the eye, whereas the dotted grey illustrates the overall $1/p$ trend. 
(b)  Similar $\gamma$ vs. $p$ for Nd-LSCO and Eu-LSCO~\cite{michon2019} and equivalent $m_\gamma$. Lines are calculated the same was as in (a), but using a different $t^\prime/t$ (see Methods~\ref{effect}). (c) $\gamma\approx S/T$ vs.$p$ of LSCO at 300~K compared with a calculation. (d)  $\gamma$ vs. $T$ of Nd-LSCO and Eu-LSCO on a semilogarithmic scale at $p=$~0.24 compared against calculated curves, using $t_z/t$ of 0.07 (solid line), and $t_z/t$ of 0 (dotted line)~(see (see Methods~\ref{effect}).  
}
\label{tailleferplot}
\end{centering}
\end{figure}

\begin{figure}[ht!]
\begin{centering}
\includegraphics[width=0.9\linewidth]{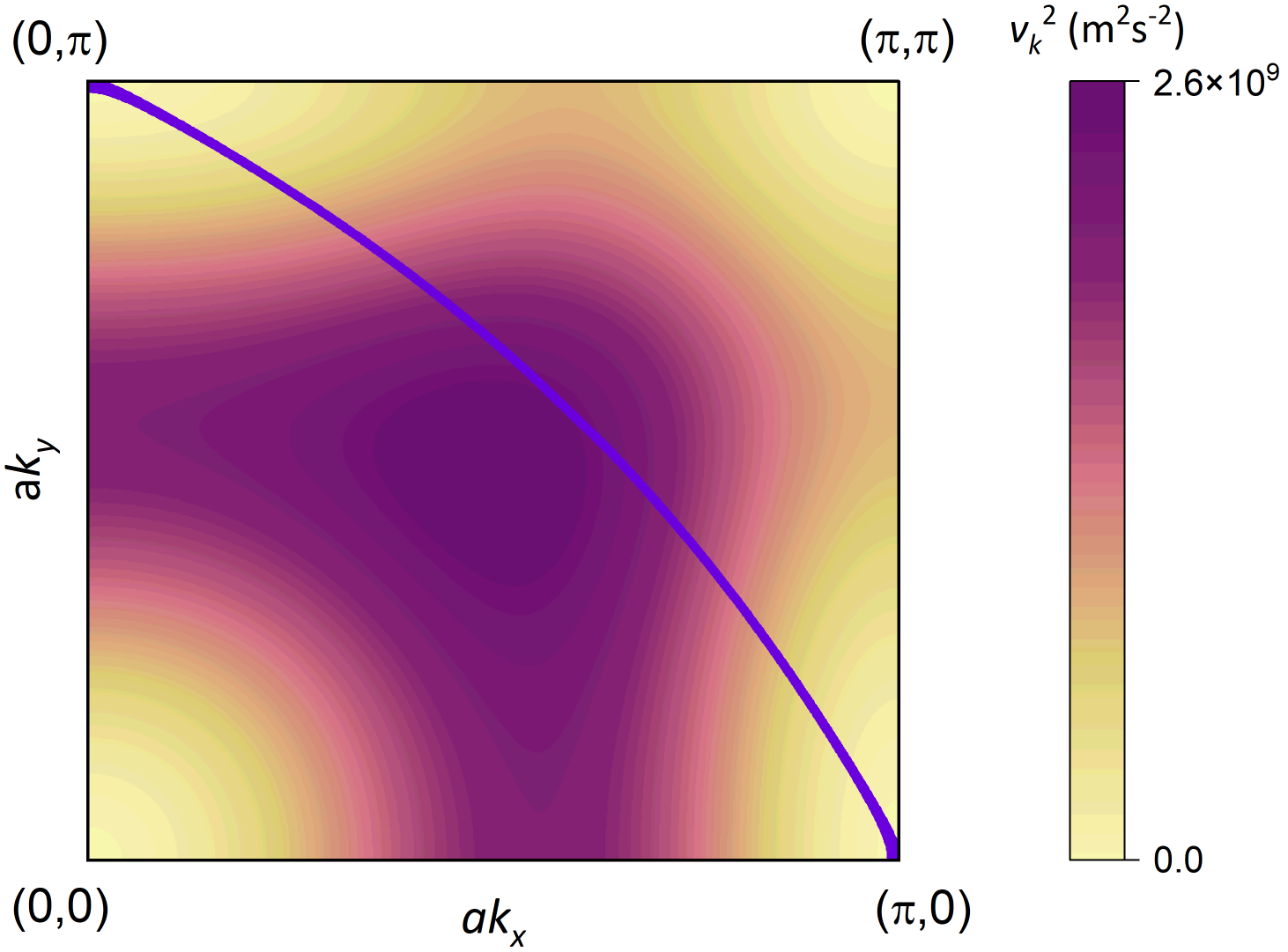}
\caption{
Momentum-space map of the velocity $v_k^2(k_x,k_y)$ over one quarter of the Brillouin zone for $p=$~0.2, together with the Fermi surface (depicted in purple). Note that because of the four-fold symmetry of the fermi surface, the Brillouin zone integration of $v_x^2$ in Methods~\ref{calculating} is equivalent to a Brillouin zone integration of $\tfrac{1}{2}v_k^2=\tfrac{1}{2}(v_x^2+v_y^2)$.
}
\label{velocitymap}
\end{centering}
\end{figure}

In Fig.~\ref{tallonplot}, we considered the effective mass $m_\gamma$ estimated from the entropy at high temperatures (and $p\gtrsim$~0.2), where it scales approximately as $1/p$. However, it has been reported that the low-temperature specific heat coefficient $\gamma(p,T)$ diverges on approaching $p \approx 0.2$~\cite{michon2019,girod2021,ramshaw2015} (see Fig.~\ref{tailleferplot}). This observation is not in contradiction, but rather reflects the fact that $\omega_{\rm opt}^2(p)$ is mostly sensitive to the high velocity nodal states whereas $\gamma(p,T)$ is  mostly sensitive to the antinodal states (see Methods~\ref{lack}). 

The insensitivity of $\omega^2_{\rm opt}(p)$ to the antinodal Van Hove singularity can be understood from the fact that it is determined by an average of the Fermi velocity over the Fermi surface, \( \omega_{\rm opt}^2 / 4\pi = e^2 \langle \kappa(k)\, v_x^2(k) \rangle \) (see Methods~\ref{calculating}). Consequently, $\omega^2_{\rm opt}(p)$ is dominated by the high-velocity nodal states on the segment of the Fermi surface intersecting the line connecting \( (0,0) \) and \( (\pi,\pi) \) in Fig.~\ref{velocitymap}, and is insensitive to the vanishing-velocity antinodal states centered near \( (0,\pi) \) and \( (\pi,0) \). Note that the grey curve (representing a non-interacting band) shows only a very small depression in $\omega_{\rm bare}^2(p)$ near \( p = 0.2 \), in sharp contrast with the singularity in \( \gamma(p) \) in Fig.~\ref{tailleferplot}. Hence, whereas $\omega_{\rm opt}^2(p)$ is determined by a Fermi-surface integral weighted by the Fermi velocity, $\gamma(p,T)$ is proportional to the density of states and thus involves an integral weighted by the inverse Fermi velocity. As a result, regions of the Fermi surface where the velocity becomes small (e.g.\ near stationary points of the dispersion) strongly enhance $\gamma(p,T)$, while contributing little to $\omega_{\rm opt}^2(p)$.

This independence can be illustrated by direct calculation of $\gamma(p,T)$ and $\omega^2_{\rm Mott}(p)$, using the effective tight-binding approximation to a Mott band. The Mott band calculation captures both the doping- and temperature-dependent entropy up to room temperature for \( p > 0.23 \) in Fig.~\ref{tallonplot}, and the low-temperature form of $\gamma(p,T)$ observed experimentally for $p\gtrsim$~0.2 [see Figs.~\ref{tailleferplot}(a) and (b)].  In LSCO, as well as in La$_{2-y-x}$Nd$_y$Sr$_x$CuO$_4$ (Nd-LSCO) and La$_{2-y-x}$Eu$_y$Sr$_x$CuO$_4$ (Eu-LSCO), the peak associated with the Van Hove singularity rides on top of a \( \gamma(p) \propto 1/p \) background, although in LSCO the full peak is preempted by the opening of the pseudogap at \( p \lesssim 0.2 \). The ability of the tight-binding approximation to a Mott band to reproduce the observed specific heat behavior for \( p \gtrsim 0.2 \) suggests that the observed feature near $p=$~0.2 is a band-structure effect rather than a quantum-critical mass divergence, as previously proposed~\cite{michon2019,girod2021}. The Van Hove singularity in the Mott band also captures all observed temperature effects, including the smooth dependence of \( \gamma(p,T) \) on doping at high temperature (i.e., at \( T = 300\,\mathrm{K} \) in Fig.~\ref{tailleferplot}(c)) and the logarithmic temperature-dependence of \( \gamma(T) \) [see Fig.~\ref{tailleferplot}(d)]. This singular behavior of $\gamma(p,T)$ contrasts sharply with $\omega_{\rm opt}^2(p)$ in Fig.~\ref{novermsimple}. This is true both for the experimental data and the model.

Note that once $p\lesssim$~0.2, $\gamma(p,T)$ is strongly suppressed by pseudogap effects, whereas $\omega_{\rm opt}^2(p)$ is not. The effect of the pseudogap on the conductivity and $\omega_{\rm opt}^2(p)$ is small because the pseudogap opens primarily in the antinodal regions of the Brillouin zone~\cite{timusk1999}, which already contribute negligibly to \( \omega_{\rm opt}^2(p) \) (see Methods~\ref{weak}).

\section*{Discussion}

We have shown that the universality of the Planckian relaxation rate follows directly from—and indeed requires—linearity of the square optical plasma frequency with doping: if the Planckian relaxation rate is universal, then \(\omega_{\rm opt}^2(p)\) must scale linearly with \(p\); conversely, if \(\omega_{\rm opt}^2(p) \propto p\), then the Planckian relaxation rate is necessarily universal;  \textcolor{black}{i.e. it is independent of doping. This universality follows from a direct comparison of the slope of the $T$-linear resistivity with the optical plasma frequency, and does not require a specific microscopic model.}
Optical-conductivity measurements have long reported such a $p$-linear doping dependence~\cite{padilla2005,legros2022,vanheumen2022,michon2021}, although little attention has been drawn to this aspect of the data. Taken together with electrical resistivity measurements, these optical conductivity measurements establish the universality of Planckian dissipation in the cuprates on purely experimental grounds.

The universality of Planckian dissipation in the cuprates raises new questions. One concerns the physical mechanism responsible for the unusually simple $p$-linear doping dependence of the square optical plasma frequency. Insight into this question can be obtained by comparison entropy measurements~\cite{loram2001,tallon2022}, which at high dopings exhibit $1/p$ scaling. The observed doping dependence of the square optical plasma frequency and the specific heat is naturally explained with  Mott physics—first applied in a similar form for liquid \(^3\)He~\cite{vollhardt1984,vollhardt1987} and later applied to the cuprates by Anderson~\cite{anderson1987}—in which the electronic dispersion is renormalized by a Gutzwiller factor \(q \approx p\). This Mott physics is distinct from, yet coexists with, the quantum-critical fluctuations believed to control Planckian dissipation. The observed electrical resistivity therefore reflects the combined effects of these two phenomena.

In contrast to Planckian dissipation, the superconducting dome is strongly doping dependent~\cite{keimer2015}. It is natural to associate the doping dependence of \(T_{\rm c}\) with the strong doping depedence of \(T_{\rm co}(p)\) (see Fig.~\ref{andoplot}), rather than with the strength of quantum-critical fluctuations behind the universal relaxation rate. The doping dependence of \(T_{\rm co}(p)\) singles out \(p^\ast\approx0.18\) as a special doping in the phase diagram~\cite{ando2004}. On the underdoped side, the physics of \(T_{\rm co}\) is associated with the pseudogap~\cite{varma1999,badoux2016}. It is noteworthy, however, that \(T_{\rm co}(p)\) collapses at $p^\ast\approx$~0.18, whereas there is evidence that the pseudogap continues to influence the thermodynamics at least up to $p\approx$~0.23~\cite{loram2001,tallon2022}. It also remains unclear what drives the strong doping dependence of \(T_{\rm co}(p)\) on the overdoped side where no pseudogap is present. 

Factoring out the square plasma frequency in the conductivity allows precision measurements of the Planckian $\alpha$. An important implication of this observed doping independence of the Planckian relaxation rate is that the scale-invariant transport of the strange metal in the cuprates is not governed by proximity to a quantum critical point at a specific doping. Rather, the dissipation mechanism appears to be intrinsic to the strange metal state itself, and need not arise from critical fluctuations associated with a particular zero-temperature ordering instability.

\acknowledgements{We thank Ross McDonald for a critical reading of the manuscript. This work was performed as part of the Department of Energy (DoE) BES project `Science of 100 tesla.' The entropy calculation procedures were developed as part of an LDRD DR project at Los Alamos National Laboratory. The National High Magnetic Field Laboratory is funded by NSF Cooperative Agreements DMR-1157490 and 1164477, the State of Florida and DoE. 
}

\section*{Data availability}
The data that support the findings of this study are available from the corresponding
author upon reasonable request.

\section*{Methods}

\appendix

\section{{Prior reports of universality}}\label{prior}

Prior reports of universal Planckian dissipation in the cuprates pertain to the regime below $T_{\rm c}$ after suppressing the superconducting state by a strong magnetic field~\cite{cooper2009,legros2019,ayres2021}. Because low-temperature resistivity is reported to exhibit the form \( \rho(T) = 1/\big((\rho_0 + A_1 T + A_2 T^2)^{-1} + \rho_{\rm max}^{-1}\big) \), as suggested in Ref.~\cite{cooper2009}, it is not scale-invariant and therefore does not display the pure $T$-linear behavior the strange metal state. Importantly, the resistivity slopes $A_1$ obtained in the low temperature regime become significantly reduced with increasing $p$~\cite{hussey2011} relative to those above $T_{\rm co}$ in Ref.~\cite{legros2019}. This contributed to the low value of $\alpha=$~0.9 at $p=$~0.26 reported by Legros {\it et al.} in Ref.~\cite{legros2019}.

We note that, more recently, it has been argued that the reported low-temperature $T$-linear resistivity in overdoped cuprates may be extrinsic, arising from disorder and from the influence of high magnetic fields on superconductivity~\cite{ramshaw2026}.

\section{{Relationship between $\omega_{\rm opt}^2$, $m^\ast$ and $n$ for anisotropic band}}\label{lack}

Our discussion of the optical conductivity has the objective to establish the value of the Planckian relaxation rate coefficient $\alpha$. As discussed in the main text, this requires only scrutinizing the measured value of the square plasma frequency in relation to the $T$-linear slope of the resistivity at high temperatures. This analysis does not require one to adopt a Drude model approximation, as is customary in prior studies of the Planckian dissipation~\cite{bruin2013,legros2019}.

Even though the Drude model breaks down for an anisotropic band, it is instructive to consider the potentially modified relationship between $\omega_{\rm opt}^2$, $n$, and $m^\ast$ in such situations. At $T=0$ and $B=0$, $\omega_{\rm opt}^2$ is derived from the Boltzmann transport equation~\cite{ashcroft1976}. Thus, $\omega_{\rm opt}^2$, $n$, and $m^\ast$ can each be expressed in terms of integrals over the Fermi surface:
\begin{align}\label{zeroplasma}
\frac{\omega_{\rm opt}^2}{4\pi e^2}
=\,&
\frac{ 1}{(2\pi)^3 \hbar}
\int_{\rm FS}
v_{\rm F}(k) \, {\rm d}S \,, \notag\\\notag\\
n
=\,&
2 \int_{\rm BZ}
\frac{d^3k}{(2\pi)^3}
\, \theta(E_F-\epsilon_k) \,, \notag\\\notag\\
m^\ast =\,& \pi \hbar^2 \, l \, D(E_F)\,,\notag\\&\hspace{1cm}\text{where}\quad
D(E_F)
=
\frac{1}{4\pi^3 \hbar}
\int_{\rm FS}
\frac{{\rm d}S}{|v_{\rm F}|}
\end{align}
is the electronic density of states~\cite{ashcroft1976}. Quite generally, these equations imply that the square optical plasma frequency can be expressed in the form
\begin{equation}\label{nonparabolicbandapproximation2}
\frac{\omega_{\rm opt}^2}{4\pi e^2}=\,C_{\epsilon_k}\!(n)\,\left(\frac{n}{m^\ast}\right), 
\end{equation}
where \(C_{\epsilon_k}\!(n)\) is a dimensionless factor that depends on the shape of the dispersion and, in general, also on the carrier density \(n\).

In the special case where the dispersion $\varepsilon_k$ is parabolic and the system possesses Galilean invariance, the above integrals become mutually dependent, such that $C_{\epsilon_k}(n)=1$. In this case, Eq.~\ref{nonparabolicbandapproximation2} reduces to the `free-space' relation, in which $\omega_{\rm opt}^2/4\pi e^2$ is expressed simply as $n/m^\ast$. A similar result holds for any isotropic dispersion, for example of the form $\epsilon_k=b\,k^s$, where $k=(k_x^2+k_y^2)^{1/2}$, $s$ is an exponent, and $b$ is a constant. In this case, the Fermi surface is circular and different definitions of the effective mass coincide, yielding again $C_{\epsilon_k}(n)=1$. This equivalence arises from the isotropy of the dispersion, which ensures that the relevant Fermi-surface averages reduce to the same quantity. There is an emergent Galilean invariance, because only the states close to the Fermi surface matter.

A qualitatively different situation occurs when the Fermi surface is anisotropic, while retaining overall four-fold symmetry. For example, for a dispersion of the form $\epsilon_k=b(|k_x|^s+|k_y|^s)$, with $s\geq 1$, which yields a diamond-shaped Fermi surface in the limit $s\to1$ and a square Fermi surface in the limit $s\to\infty$, one obtains
$C_{\epsilon_k}\!(n)=({s^2}/{(s - 1)}) \sin\!\left({\pi}/{s}\right)/4$.
This is a fixed dimensionless factor between $\pi/4$ and 1. It reaches its maximum value, \(C_{\epsilon_k}=1\), for the parabolic band with \(s=2\), and is less than 1 for an anisotropic band given by \(s\neq 2\).

Another, even simpler, example of four-fold anisotropy is a Fermi surface consisting of two elliptical pockets with orthogonal major axes. In this case, one obtains
$C_{\epsilon_k}(n)={2r}/(1+r^2)$,
where \(r\) is the ratio of the major to minor axes of the ellipse. Here, \(0 \leq C_{\epsilon_k} \leq 1\), with \(C_{\epsilon_k} \to 0\) in the limits of vanishing or large \(r\). Note that \(C_{\epsilon_k}\) again depends on the shape of the dispersion, which in this case is characterized by \(r\).

Finally, a more realistic example is provided by a tight-binding dispersion. For illustrative purposes, we consider the simplest case of a square-lattice tight-binding band, $\epsilon_k=2t[\cos(ak_x)+\cos(ak_y)]$. In this case, \(C_{\epsilon_k}(n)\) depends explicitly on the chemical potential \(\mu\), or equivalently on the carrier density \(n\). In the limit \(n\to 0\), the bottom of the band is approximately parabolic, such that \(C_{\epsilon_k}(n\to0)\to 1\). A similar result is obtained near the top of the band, where \(n\to 2/a^2\), and the dispersion again becomes locally parabolic. Since \(n=(1+p)/a^2\), it follows that \(C_{\epsilon_k}(n\to0~{\rm or}~2/a^2)\to 1\) in the limits \(p\to \pm 1\). This behavior underlies the approximately linear dependence of \(\omega_{\rm bare}^2\) on \(p\) in Fig.~\ref{novermsimple}(a), although the dispersion of LSCO includes additional hopping terms beyond nearest neighbors~\cite{horio2018}.

While the tight-binding dispersion is approximately parabolic near the top and bottom of the band, the hole dopings relevant to superconductivity occur near the middle of the band, where a van Hove singularity is present. For the simple tight-binding model considered above, this singularity occurs at \(p=0\), whereas in LSCO it occurs near \(p\approx 0.2\)~\cite{horio2018}. Because \(\omega_{\rm bare}^2\) remains finite at the van Hove singularity (see Fig.~\ref{novermsimple}(a)), whereas \(m^\ast\) diverges (see Fig.~\ref{tailleferplot}), it follows that \(C_{\epsilon_k}(n)\) must also become singular and strongly dependent on \(n\). It is therefore clear that \(\omega_{\rm opt}^2\) cannot be expressed as a function of \(n/m^\ast\) alone in a way that is useful. 

One commonly used convention is to define an optical mass so as to preserve the appearance of a Galilean-invariant relation, $\omega_{\rm opt}^2/4\pi e^2 = n/m_{\rm opt}$. Values of \(m_{\rm opt}\) obtained in this way—assuming \(n=1/eR_{\rm H}\)~\cite{ando2004}—by Padilla {\it et al.}~\cite{padilla2005} are included in Fig.~\ref{tailleferplot}(a) (denoted by \(+\) symbols). Such a definition corresponds to \(m_{\rm opt}=m^\ast/C_{\epsilon_k}(n)\). This makes it clear that for a tight-binding dispersion, \(m_{\rm opt}\) and \(m^\ast\) will always exhibit different doping dependences. A direct comparison of these quantities therefore requires the adoption of a specific model. The definition of \(m_{\rm opt}\) is, in effect, tautological, since it simply provides an alternative way of expressing \(\omega_{\rm opt}^2\) and does not constitute an independently measurable quantity. By contrast, cyclotron mass measurements~\cite{legros2022} (\(\times\) symbols in Fig.~\ref{tailleferplot}(a)) break this tautology, as the mass is determined independently of \(\omega_{\rm opt}^2\).

\section{{Calculating the square plasma frequency}}\label{calculating} 


The squared plasma frequency of the Mott band is calculated using~\cite{ashcroft1976}
\begin{align}
\frac{\omega_{\rm Mott}^2}{4\pi e^2}
= \big\langle \nu(k)\, v_x^2(k) \big\rangle\nonumber\hspace{4.1cm}\\
= 2 \int_{\rm BZ} \frac{{\rm d}^3k}{(2\pi)^3}\,
v_x^2(k)\,\Bigl(-\frac{\partial n_{\rm F}(E,T)}{\partial E}\Bigr)\Big|_{E=\epsilon_k},
\label{velocities}
\end{align}
where \( v_x = \hbar^{-1}\partial\epsilon_k/\partial k_x \) is the group velocity along the $x$-direction (only a single direction being required in a four-fold symmetric system), and $\varepsilon_k$ is the tight binding approximation, including the prefactor of $q$ given by Eq.~\ref{massrenormalization}. \textcolor{black}{In the zero-temperature limit, this expression can be reduced to the form
%
\begin{equation}\label{zeroMottband}
\frac{\omega_{\rm Mott}^2}{4\pi e^2}
=
\frac{1}{(2\pi)^3 \hbar}
\int_{\rm FS}
v_{\rm F}(k) \, {\rm d}S ,
\end{equation}
%
which is an integral over the Fermi surface area~\cite{ashcroft1976,timusk1999}.} Because the observed plasma frequency is given by $\omega_{\rm opt}^2 / 4\pi=e^2 \langle \kappa(k) v_x^2(k) \rangle$, an additional factor of $(1+F_0^{\rm s})^{-1}$ is required to bring the results of Eq.~(\ref{velocities}) into alignment with the experimental values of $\omega_{\rm opt}^2(p)$, as described in the main text. 

For a non-interacting band [grey curve in Fig.~\ref{novermsimple}(a)], we set $q=1$, in which case Eq.~(\ref{velocities}) corresponds to the optical sum-rule of that band. The resulting values for the square plasma frequency $\omega^2_{\rm bare}(p)$ of a non-interacting bare band are plotted in Fig.~\ref{novermsimple}(a).


\section{{Linear regression to determine \(A\) and~\(T_{\rm co}\)}}\label{linear}

\begin{figure}[ht!]
\centering
\includegraphics[width=0.9\linewidth]{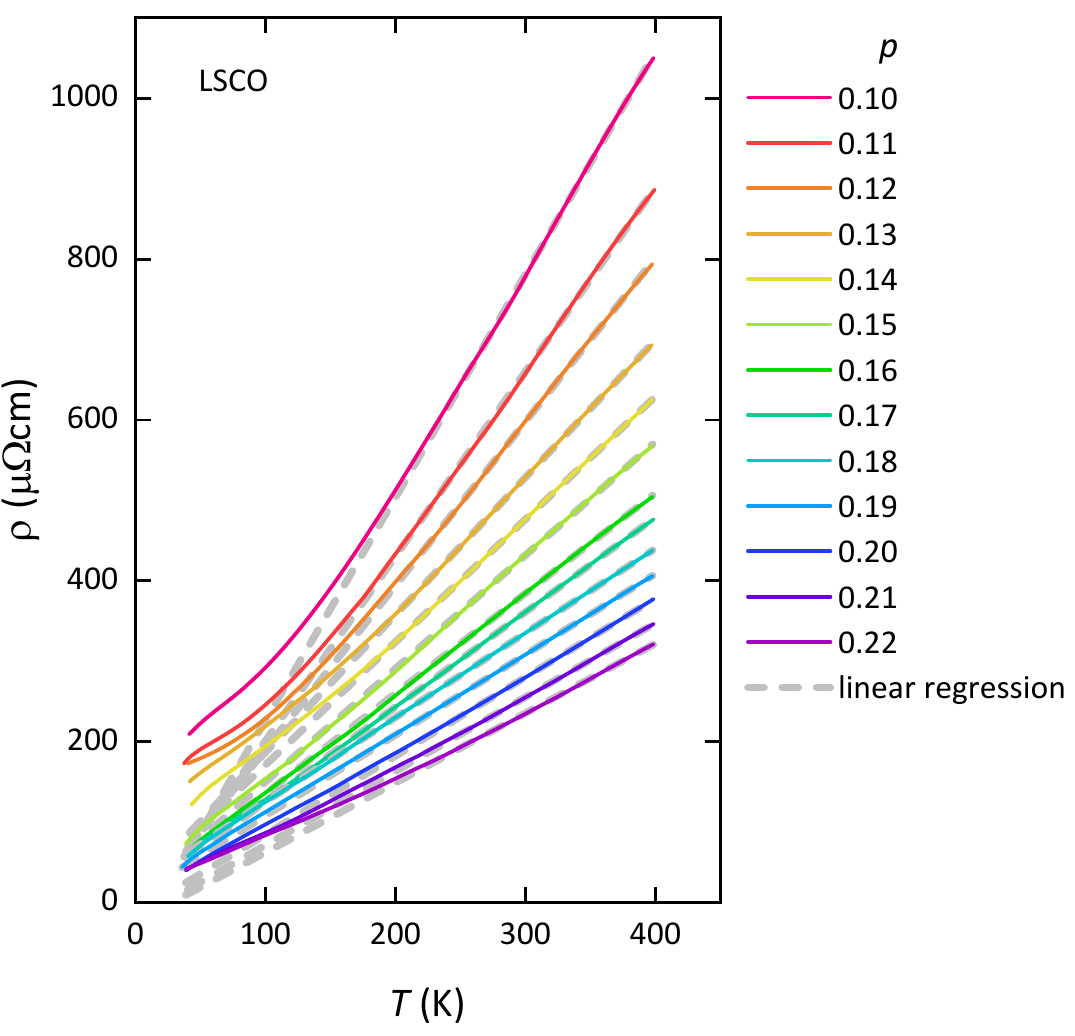}
\caption{
Measured $\rho$ versus $T$ at different hole dopings~\cite{ando2004}, as indicated in different colors. Dashed lines indicate fits to determine $A$. The obtained values of $1/A$ are plotted in Fig.~\ref{andoplot}(b). 
}
\label{andofits}
\end{figure}

To establish whether the data is well described by a \(T\)-linear behavior at high temperatures in Fig.~\ref{andoplot}(a), in Fig.~\ref{andofits} we fit the function \(\rho=\rho_0+AT\) to the in-plane resistivity data in LSCO~\cite{ando2004} above 250~K for each hole doping. While this temperature is chosen arbitrarily, we find that the linear regression continues to describe the data for a range of temperatures below this value. At \(p^\ast = 0.18\)~\cite{ando2004}, the \(T\)-linear behavior extends all the way from 400~K down to \(T_{\rm c}\). We define the crossover temperature \(T_{\rm co}\) as the characteristic temperature below which the resistivity deviates from linearity. We choose 3~$\mu\Omega$cm as the threshold above which a deviation from linearity is considered significant and estimate the error in \(T_{\rm co}\) by increasing the threshold to 10~$\mu\Omega$cm.

The downward departure of \(1/A\) from the dashed line below \(p \approx 0.15\) in Fig.~\ref{andoplot}(a) could arise either from a larger effect of the pseudogap at lower dopings or from a larger crossover temperature at these dopings. It is also well-known that at these lower dopings, disorder effects are prevalent~\cite{rullieralbenque2008}, leading to a glassy behavior approaching \(p \sim 0.05\).


\section{{Tight-binding approximation of a Mott band}}\label{tight} 

Throughout, we assume the tight-binding approximation for a Mott band, where the electronic dispersion is renormalized by a factor $q$~\cite{brinkman1970,kotliar1988} so that
\begin{align}\label{tightbindingdispersion}
\epsilon_k=q\,\Big[-2t\big(\cos(ak_x)+\cos(ak_y)\big)\hspace{3cm}\nonumber\\
\hspace{0.3cm}+~4t^\prime\cos(ak_x)\cos (ak_y)-2t^{\prime\prime}\big(\cos(2ak_x)+\cos(2ak_y)\big)\nonumber\\
+~4t^{\prime\prime\prime}\big(\cos(ak_x)\cos(2ak_y)+\cos(ak_y)\cos(2ak_x)\big)\hspace{0.75cm}\nonumber\\
+~2t_z\cos(ak_x/2)\cos(ak_y/2)\big[\cos(ak_x)-\cos(ak_y)\big]^2\hspace{0.35cm}\nonumber\\
\times\cos(ck_z)\Big]+\mu.\hspace{5.6cm}
\end{align}
Here, \(k_x\), \(k_y\), and \(k_z\) represent components of the reciprocal lattice vector, \(a\approx\)~3.8~\AA~is the planar lattice spacing, \(c=2l\approx\)~13.2~\AA~is the interlayer lattice spacing, and \(\mu\) is the chemical potential, adjusted to produce an unreconstructed Fermi surface cross-section comprising \(1+p\) holes for $p<$~0.2 or $1-p$ electrons for $p>$~0.2, consistent with Luttinger's theorem~\cite{luttinger1960a,luttinger1960b}. We use the nearest neighbor hopping, \(t = 430~\text{meV}\), as determined for the bare conduction band by electronic structure calculations~\cite{pavarini2001,mattheiss1987,singh1994}, and ratios \(t^\prime/t\), \(t^{\prime\prime}/t^\prime\), and \(t_z/t\) that are fixed. These ratios have been determined elsewhere~\cite{horio2018}: 
\begin{align}
\begin{array}{l|cccc} 
  & t^\prime/t & t^{\prime\prime}/t^\prime & t^{\prime\prime\prime}/t^{\prime\prime} & t_z/t \\
 \hline 
\mbox{LSCO} & 0.12 & 0.50 & 0 & 0.07 \\
\mbox{Nd-LSCO} & 0.14 & 0.50 & 0 & 0.07 \\ 
\mbox{Eu-LSCO} & 0.14 & 0.50 & 0 & 0.07 \\ 
\end{array}\:\:.
\end{align}

The electronic density of states per copper oxide plane is~\cite{ashcroft1976}
\begin{align}\label{spectralfunction}
\rho(E) = {2la^2}\int\frac{{\rm d}^3k}{(2\pi)^3} \; \delta\big(\epsilon_k - E\big) \,.
\end{align}
Here, the prefactor of 2 corresponds to the number of spins, and \(\delta\big(\epsilon_k - E\big)\) is a delta function defined such that \(\int \rho(E){\rm d}E=2\). The entropy is then given by~\cite{ashcroft1976}:
\begin{equation}\label{entropyintegral}
S  =  N_{\rm A}\int\limits_{-\infty}^{\infty}\, 
	{\rm d}E \, \rho(E) \,\big[n_{\rm F}\ln n_{\rm F}+[1-n_{\rm F}] \ln(1-n_{\rm F})\big]   \,,
\end{equation} 
where \(n_{\rm F}=1/(1+\exp\{E/k_{\rm B}T\})\) is the Fermi-Dirac distribution, and \(N_{\rm A}\) is Avogadro's number. 

It is important to note here that while $q$ represents the overall renormalization of the band, part of Fermi surface is subsequently gapped (by the pseudogap) at $p<0.23$, leading to a net suppression of $m^\ast$ (and $m_\gamma$) at low temperatures, as observed in Fig.~\ref{tailleferplot}(a), (b) and (c). In such a case, the unrenormalized band mass $m=qm^\ast$ must also be considered as a quantity affected by the gap at $p<0.23$ so as to ensure that $q=m/m^\ast$ throughout.

\section{{Determining $q$ from the electronic entropy}}\label{determining}

\begin{figure}[ht!]
\begin{centering}
\includegraphics[width=0.9\linewidth]{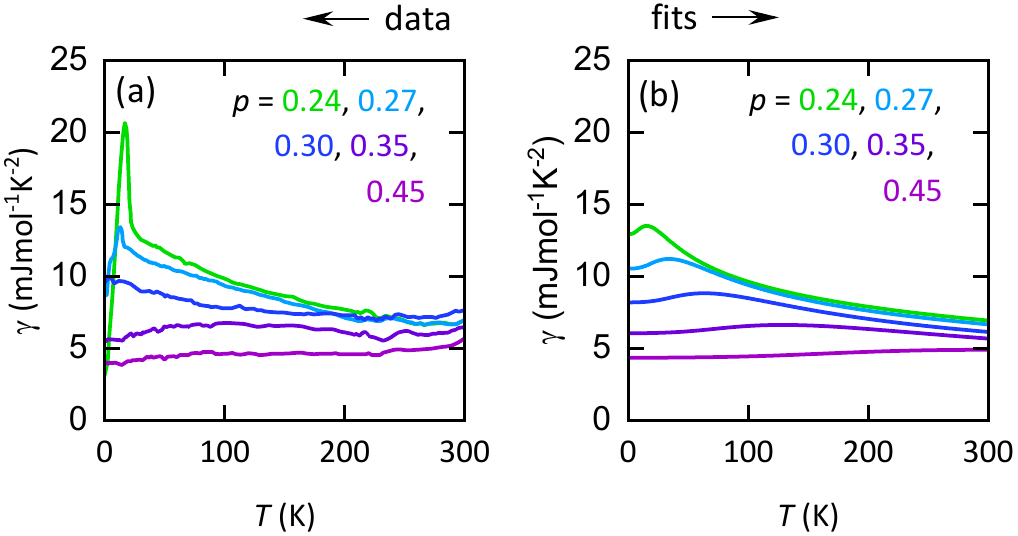}
\caption{
(a) Measured electronic coefficient $\gamma$ of the specific heat. (b) Calculated $\gamma$ versus $T$ curves using fitted values of $m/m^\ast$. 
}
\label{tallongamma}
\end{centering}
\end{figure}

Using the above fixed values of \(t\) and the ratios \(t^\prime/t\), \(t^{\prime\prime}/t^\prime\), and \(t_z/t\), we calculate the entropy in Fig.~\ref{tallonplot}(c) using Eq.~\ref{entropyintegral}. Agreement with the experimentally measured entropy in Fig.~\ref{tallonplot}(a) is achieved by adjusting \(q\) in Eq.~\ref{tightbindingdispersion}. The resulting values of \(q\) are plotted in Fig.~\ref{tallonplot}(d). Figure~\ref{tallongamma} compares the measured temperature dependence of $\gamma$ against that calculated. Note the presence of a superconducting anomaly for $p\leq$~0.27, which is not included in the tight-binding model,

\begin{figure}[ht!]
\begin{centering}
\includegraphics[width=0.9\linewidth]{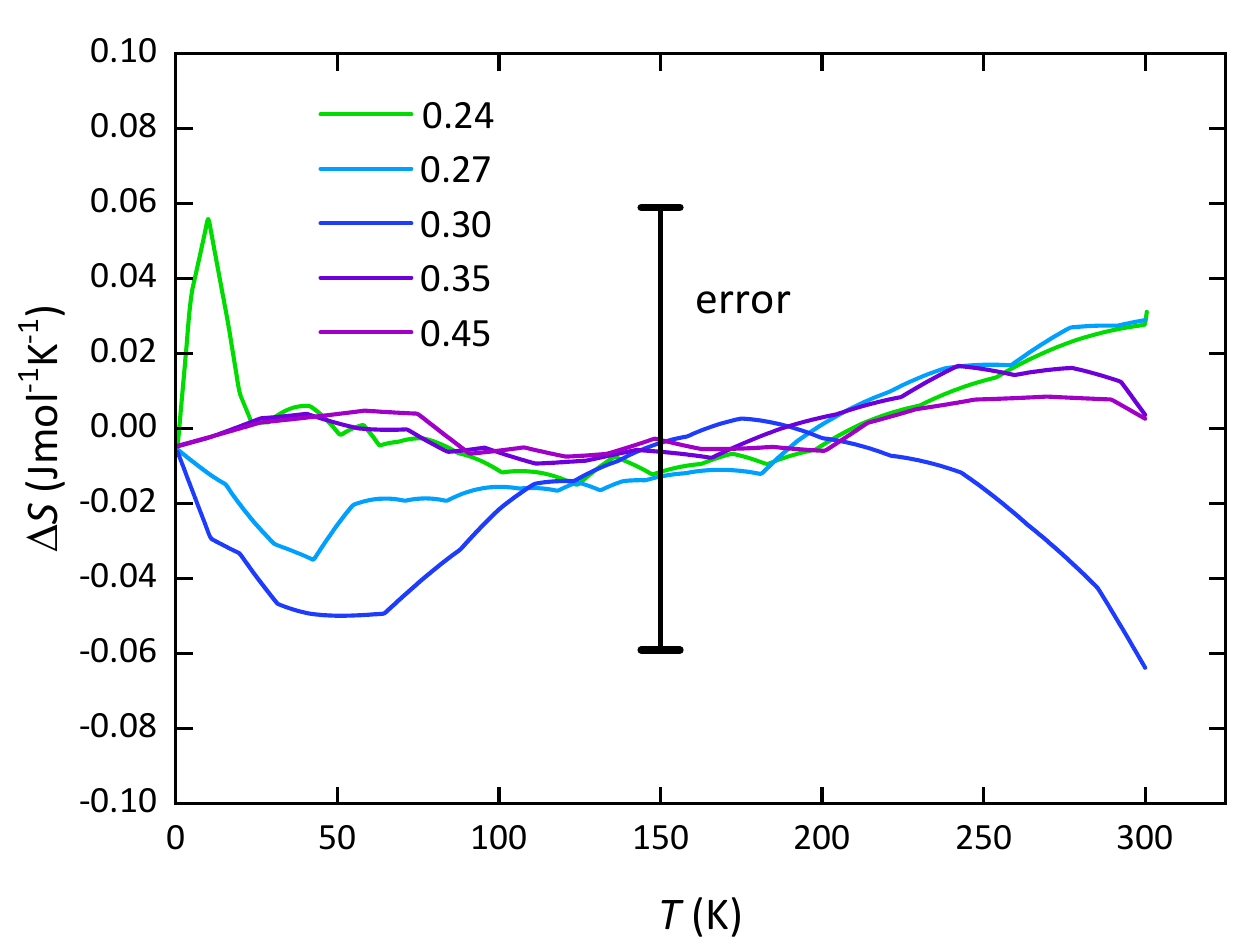}
\caption{
Difference $\Delta S$ between the measured and calculated entropy in Figs~3(a) and (c) of the main text. The different colors refer to the different hole dopings. 
}
\label{entropyerror}
\end{centering}
\end{figure}

The errors (see Fig.~\ref{entropyerror}) are estimated from the small difference $\Delta S$ of the calculated entropy curve from the experimental curve. This shows no systematic trends, suggesting that the difference is primarily the result of random error. We assume the largest deviation of $\Delta S$ from zero to provide an estimate of the error $\delta S$ in $S$. Since the magnitude of the entropy at a given temperature scales roughly with the effective mass, we can infer an approximate error for the fitted effective mass given by $\delta q=(\delta S/S)q$. 

\section{{Effect of interlayer hopping on the Van Hove singularity}}\label{effect}

At temperatures below $\sim~$2~K, \(\gamma\) in Fig.~\ref{tailleferplot}(d) becomes sensitive to the interlayer hopping. Whereas Ref.~\cite{horio2018} finds \(t_z/t = 0.07\), Ref.~\cite{zhong2022} finds \(t_z/t = 0.03\). The measurements of \(\gamma\) close to \(p = 0.23\) were performed under strong magnetic fields~\cite{michon2019}. It has been argued that orbital-averaging leads to a suppression of the interlayer hopping~\cite{musser2022}, which could cause the effective ratio \(t_z/t\) to drop to a significantly smaller value. For \(t_z/t = 0\), the logarithmic divergence in Fig.~\ref{tailleferplot}(d) continues down to \(T = 0\).

Impurity scattering introduces a residual scattering rate at \(T = 0\), which can also cause \(\gamma\) to saturate~\cite{michon2019}. Angle-dependent magnetoresistance measurements have found the residual scattering rate to be momentum-dependent~\cite{grissonnanche2021}. However, since the Fermi velocity vanishes along all momentum directions at the point of the Fermi surface where the Van Hove singularity occurs, angle-dependent magnetoresistance measurements are unable to constrain a value for this scattering rate at this point on the Fermi surface~\cite{grissonnanche2021}. In order for the mean free path to remain non-vanishing at this point, \(\tau^{-1}\) must vanish at this point.

\section{{Weak effect of the pseudogap on the conductivity}}\label{weak}

The opening of the pseudogap has very little impact on \( \omega_{\rm opt}^2(p) \) in LSCO. This contrasts with the strong reduction in $\gamma(p,T)$ observed for $p\lesssim$~0.2 in Fig.~\ref{tailleferplot}(a) \& (b)~\cite{loram2001,momono2002,girod2021,michon2019}. 

Experimental evidence for the weak sensitivity of the plasma frequency squared to the pseudogap includes: (i) the absence of any discernible jump in \( \omega_{\rm opt}^2(p) \) across \( p^\ast\approx0.18 \) in Figs.~\ref{drudeweight} and \ref{novermsimple}(b); (ii) the weak temperature dependence of \( \omega_{\rm opt}^2(p,T) \) observed in optical conductivity measurements on Bi2201~\cite{vanheumen2022} and on Bi\(_2\)Sr\(_2\)Ca\(_{0.92}\)Y\(_{0.08}\)Cu\(_2\)O\(_{8-\delta}\) (Bi2212)~\cite{zaanen2019,vandermarel2003}; and (iii) a tight-binding calculation for a Mott-band (see Methods~\ref{calculating} and~\ref{tight}) that quantitatively reproduces—using the Landau compressibility interaction factor—the values of \( \omega_{\rm opt}^2(p) \) obtained from optical conductivity, THz spectroscopy, and penetration depth measurements [Fig.~\ref{novermsimple}(b), purple line] via Eq.~\ref{massrenormalization}, without introducing any additional parameters and, in particular, without including the pseudogap.

Because the square optical plasma frequency is proportional to the Fermi surface average of the velocity,  it is predominantly weighted by states with large Fermi velocity, i.e., the nodal regions. The absence of any discernible change in the measured $\omega_{\rm opt}^2$ across $p^\ast$ therefore indicates that any proposed Fermi surface reconstruction occurring for $p < p^\ast$ cannot significantly affect the nodal spectrum. It is interesting that photoemission measurements on LSCO~\cite{yoshida2006} also show that the dispersion near the nodal directions remains essentially unchanged across $p^\ast$.




\bibliographystyle{naturemag}


\begin{thebibliography}{10}
\expandafter\ifx\csname url\endcsname\relax
  \def\url#1{\texttt{#1}}\fi
\expandafter\ifx\csname urlprefix\endcsname\relax\def\urlprefix{URL }\fi
\providecommand{\bibinfo}[2]{#2}
\providecommand{\eprint}[2][]{\url{#2}}

\bibitem{bednortz1986}
\bibinfo{author}{Bednortz, J.~G.} \& \bibinfo{author}{Muller, K.~A.}
\newblock \bibinfo{title}{Possible high $t_c$ superconductivity in the
  ba-la-cu-o system}.
\newblock \emph{\bibinfo{journal}{Z. Phys.}} \textbf{\bibinfo{volume}{64}},
  \bibinfo{pages}{189--193} (\bibinfo{year}{1986}).

\bibitem{anderson1988}
\bibinfo{author}{Anderson, P.~W.}
\newblock \bibinfo{title}{Strange insulators, strange semiconductors, strange
  metals: High tc as a case history in condensed-matter physics}.
\newblock In \emph{\bibinfo{booktitle}{AIP Conference Proceedings}}, vol.
  \bibinfo{volume}{169}, \bibinfo{pages}{141--157} (\bibinfo{publisher}{AIP
  Publishing}, \bibinfo{year}{1988}).

\bibitem{landau1957}
\bibinfo{author}{Landau, L.~D.}
\newblock \bibinfo{title}{The theory of a fermi liquid}.
\newblock \emph{\bibinfo{journal}{Sov. Phys. JETP}}
  \textbf{\bibinfo{volume}{3}}, \bibinfo{pages}{920} (\bibinfo{year}{1957}).

\bibitem{anderson1992}
\bibinfo{author}{Anderson, P.~W.}
\newblock \bibinfo{title}{Experimental constraints on the theory of high-tc
  superconductivity}.
\newblock \emph{\bibinfo{journal}{Science}} \textbf{\bibinfo{volume}{256}},
  \bibinfo{pages}{1526--1531} (\bibinfo{year}{1992}).

\bibitem{zaanen2004}
\bibinfo{author}{Zaanen, J.}
\newblock \bibinfo{title}{Why the temperature is high}.
\newblock \emph{\bibinfo{journal}{Nature}} \textbf{\bibinfo{volume}{430}},
  \bibinfo{pages}{512--513} (\bibinfo{year}{2004}).

\bibitem{keimer2015}
\bibinfo{author}{Keimer, B.}, \bibinfo{author}{Kivelson, S.~A.},
  \bibinfo{author}{Norman, M.~R.}, \bibinfo{author}{Uchida, S.} \&
  \bibinfo{author}{Zaanen, J.}
\newblock \bibinfo{title}{From quantum matter to high-temperature
  superconductivity in copper oxides}.
\newblock \emph{\bibinfo{journal}{Nature}} \textbf{\bibinfo{volume}{518}},
  \bibinfo{pages}{179--186} (\bibinfo{year}{2015}).

\bibitem{PinesNozieres1999}
\bibinfo{author}{Pines, D.} \& \bibinfo{author}{Nozieres, P.}
\newblock \emph{\bibinfo{title}{Theory Of Quantum Liquids}}
  (\bibinfo{publisher}{CRC Press}, \bibinfo{year}{1999}).

\bibitem{bruin2013}
\bibinfo{author}{Bruin, J. A.~N.}, \bibinfo{author}{Sakai, H.},
  \bibinfo{author}{Perry, R.~S.} \& \bibinfo{author}{Mackenzie, A.~P.}
\newblock \bibinfo{title}{Similarity of scattering rates in metals showing
  t-linear resistivity}.
\newblock \emph{\bibinfo{journal}{Science}} \textbf{\bibinfo{volume}{339}},
  \bibinfo{pages}{804--807} (\bibinfo{year}{2013}).

\bibitem{hartnoll2022}
\bibinfo{author}{Hartnoll, S.~A.} \& \bibinfo{author}{Mackenzie, A.~P.}
\newblock \bibinfo{title}{<i>colloquium:</i> planckian dissipation in metals}.
\newblock \emph{\bibinfo{journal}{Reviews of Modern Physics}}
  \textbf{\bibinfo{volume}{94}} (\bibinfo{year}{2022}).

\bibitem{phillips2022}
\bibinfo{author}{Phillips, P.~W.}, \bibinfo{author}{Hussey, N.~E.} \&
  \bibinfo{author}{Abbamonte, P.}
\newblock \bibinfo{title}{Stranger than metals}.
\newblock \emph{\bibinfo{journal}{Science}} \textbf{\bibinfo{volume}{377}}
  (\bibinfo{year}{2022}).
\newblock \urlprefix\url{https://doi.org/10.1126/science.abh4273}.

\bibitem{cao2020}
\bibinfo{author}{Cao, Y.} \emph{et~al.}
\newblock \bibinfo{title}{Strange metal in magic-angle graphene with near
  planckian dissipation}.
\newblock \emph{\bibinfo{journal}{Phys. Rev. Lett.}}
  \textbf{\bibinfo{volume}{124}}, \bibinfo{pages}{076801}
  (\bibinfo{year}{2020}).
\newblock
  \urlprefix\url{https://journals.aps.org/prl/abstract/10.1103/PhysRevLett.124.076801}.

\bibitem{Varma1989}
\bibinfo{author}{Varma, C.~M.}, \bibinfo{author}{Littlewood, P.~B.},
  \bibinfo{author}{Schmitt-Rink, S.}, \bibinfo{author}{Abrahams, E.} \&
  \bibinfo{author}{Ruckenstein, A.}
\newblock \bibinfo{title}{Phenomenology of the normal state of cu-o
  high-temperature superconductors}.
\newblock \emph{\bibinfo{journal}{Phys. Rev. Lett.}}
  \textbf{\bibinfo{volume}{63}}, \bibinfo{pages}{1996} (\bibinfo{year}{1989}).

\bibitem{nagaosa1992}
\bibinfo{author}{Nagaosa, N.}
\newblock \bibinfo{title}{Rvb vs fermi liquid picture of high-tc
  superconductors}.
\newblock \emph{\bibinfo{journal}{Journal of Physics and Chemistry of Solids}}
  \textbf{\bibinfo{volume}{53}}, \bibinfo{pages}{1493--1498}
  (\bibinfo{year}{1992}).

\bibitem{martin1990}
\bibinfo{author}{Martin, S.}, \bibinfo{author}{Fiory, A.~T.},
  \bibinfo{author}{Fleming, R.~M.}, \bibinfo{author}{Schneemeyer, L.~F.} \&
  \bibinfo{author}{Waszczak, J.~V.}
\newblock \bibinfo{title}{Normal-state transport properties of biscco
  crystals}.
\newblock \emph{\bibinfo{journal}{Phys. Rev. B}} \textbf{\bibinfo{volume}{41}},
  \bibinfo{pages}{846--849} (\bibinfo{year}{1990}).
\newblock \urlprefix\url{https://link.aps.org/doi/10.1103/PhysRevB.41.846}.

\bibitem{ando2004}
\bibinfo{author}{Ando, Y.}, \bibinfo{author}{Komiya, S.},
  \bibinfo{author}{Segawa, K.}, \bibinfo{author}{Ono, S.} \&
  \bibinfo{author}{Kurita, Y.}
\newblock \bibinfo{title}{Electronic phase diagram of high-tc cuprate
  superconductors from a mapping of the in-plane resistivity curvature}.
\newblock \emph{\bibinfo{journal}{Phys. Rev. Lett.}}
  \textbf{\bibinfo{volume}{93}}, \bibinfo{pages}{267001}
  (\bibinfo{year}{2004}).
\newblock
  \urlprefix\url{https://link.aps.org/doi/10.1103/PhysRevLett.93.267001}.

\bibitem{wilson1974}
\bibinfo{author}{Wilson, K.~G.} \& \bibinfo{author}{Kogut, J.}
\newblock \bibinfo{title}{The renormalization group and the
  {$\epsilon$}-expansion}.
\newblock \emph{\bibinfo{journal}{Physics Reports}}
  \textbf{\bibinfo{volume}{12}}, \bibinfo{pages}{75--199}
  (\bibinfo{year}{1974}).

\bibitem{lohneysen2007}
\bibinfo{author}{{v.~L\"ohneysen}, H.}, \bibinfo{author}{Rosch, A.},
  \bibinfo{author}{Vojta, M.} \& \bibinfo{author}{W\"olfle, P.}
\newblock \bibinfo{title}{Fermi-liquid instabilities at magnetic quantum phase
  transitions}.
\newblock \emph{\bibinfo{journal}{Reviews of Modern Physics}}
  \textbf{\bibinfo{volume}{79}}, \bibinfo{pages}{1015--1075}
  (\bibinfo{year}{2007}).

\bibitem{emery1995i}
\bibinfo{author}{Emery, V.~J.} \& \bibinfo{author}{Kivelson, S.~A.}
\newblock \bibinfo{title}{Superconductivity in bad metals}.
\newblock \emph{\bibinfo{journal}{Physical Review Letters}}
  \textbf{\bibinfo{volume}{74}}, \bibinfo{pages}{3253--3256}
  (\bibinfo{year}{1995}).

\bibitem{zaanen2019}
\bibinfo{author}{Zaanen, J.}
\newblock \bibinfo{title}{Planckian dissipation, minimal viscosity and the
  transport in cuprate strange metals}.
\newblock \emph{\bibinfo{journal}{SciPost Phys.}} \textbf{\bibinfo{volume}{6}},
  \bibinfo{pages}{061} (\bibinfo{year}{2019}).
\newblock \urlprefix\url{https://scipost.org/10.21468/SciPostPhys.6.5.061}.

\bibitem{varma2020}
\bibinfo{author}{Varma, C.~M.}
\newblock \bibinfo{title}{Colloquium: Linear in temperature resistivity and
  associated mysteries including high temperature superconductivity}.
\newblock \emph{\bibinfo{journal}{Rev. Mod. Phys.}}
  \textbf{\bibinfo{volume}{92}}, \bibinfo{pages}{031001}
  (\bibinfo{year}{2020}).
\newblock
  \urlprefix\url{https://link.aps.org/doi/10.1103/RevModPhys.92.031001}.

\bibitem{davison2014}
\bibinfo{author}{Davison, R.~A.}, \bibinfo{author}{Schalm, K.} \&
  \bibinfo{author}{Zaanen, J.}
\newblock \bibinfo{title}{Holographic duality and the resistivity of strange
  metals}.
\newblock \emph{\bibinfo{journal}{Phys. Rev. B}} \textbf{\bibinfo{volume}{89}},
  \bibinfo{pages}{245116} (\bibinfo{year}{2014}).
\newblock
  \urlprefix\url{https://journals.aps.org/prb/abstract/10.1103/PhysRevB.89.245116}.

\bibitem{aji2007}
\bibinfo{author}{Aji, V.} \& \bibinfo{author}{Varma, C.~M.}
\newblock \bibinfo{title}{Theory of the quantum critical fluctuations in
  cuprate superconductors}.
\newblock \emph{\bibinfo{journal}{Phys. Rev. Lett.}}
  \textbf{\bibinfo{volume}{99}}, \bibinfo{pages}{067003}
  (\bibinfo{year}{2007}).
\newblock
  \urlprefix\url{https://journals.aps.org/prl/abstract/10.1103/PhysRevLett.99.067003}.

\bibitem{patel2023}
\bibinfo{author}{Patel, A.~A.}, \bibinfo{author}{Guo, H.},
  \bibinfo{author}{Esterlis, I.} \& \bibinfo{author}{Sachdev, S.}
\newblock \emph{\bibinfo{journal}{Science}} \textbf{\bibinfo{volume}{381}},
  \bibinfo{pages}{790--793} (\bibinfo{year}{2023}).

\bibitem{chang2025}
\bibinfo{author}{Chang, Y.-Y.}, \bibinfo{author}{Nguyen, K.~V.},
  \bibinfo{author}{Remund, K.} \& \bibinfo{author}{Chung, C.-H.}
\newblock \bibinfo{title}{Rep.~prog.~phys.}
\newblock \emph{\bibinfo{journal}{Reports on Progress in Physics}}
  \textbf{\bibinfo{volume}{88}}, \bibinfo{pages}{048001}
  (\bibinfo{year}{2025}).

\bibitem{legros2019}
\bibinfo{author}{Legros, A.} \emph{et~al.}
\newblock \bibinfo{title}{Universal $t$-linear resistivity and planckian
  dissipation in overdoped cuprates}.
\newblock \emph{\bibinfo{journal}{Nature Physics}}
  \textbf{\bibinfo{volume}{15}}, \bibinfo{pages}{142--147}
  (\bibinfo{year}{2019}).

\bibitem{takagi1992}
\bibinfo{author}{Takagi, H.} \emph{et~al.}
\newblock \bibinfo{title}{Systematic evolution of temperature-dependent
  resistivity in lsco}.
\newblock \emph{\bibinfo{journal}{Phys. Rev. Lett.}}
  \textbf{\bibinfo{volume}{69}}, \bibinfo{pages}{2975--2978}
  (\bibinfo{year}{1992}).
\newblock \urlprefix\url{https://link.aps.org/doi/10.1103/PhysRevLett.69.2975}.

\bibitem{ito1993}
\bibinfo{author}{Ito, T.}, \bibinfo{author}{Takenaka, K.} \&
  \bibinfo{author}{Uchida, S.}
\newblock \bibinfo{title}{Systematic deviation from t-linear behavior in the
  in-plane resistivity of ybco: Evidence for dominant spin scattering}.
\newblock \emph{\bibinfo{journal}{Phys. Rev. Lett.}}
  \textbf{\bibinfo{volume}{70}}, \bibinfo{pages}{3995--3998}
  (\bibinfo{year}{1993}).
\newblock \urlprefix\url{https://link.aps.org/doi/10.1103/PhysRevLett.70.3995}.

\bibitem{watanabe1997}
\bibinfo{author}{Watanabe, T.}, \bibinfo{author}{Fujii, T.} \&
  \bibinfo{author}{Matsuda, A.}
\newblock \bibinfo{title}{Anisotropic resistivities of precisely oxygen
  controlled single-crystal bscco: Systematic study on ``spin gap'' effect}.
\newblock \emph{\bibinfo{journal}{Phys. Rev. Lett.}}
  \textbf{\bibinfo{volume}{79}}, \bibinfo{pages}{2113--2116}
  (\bibinfo{year}{1997}).
\newblock \urlprefix\url{https://link.aps.org/doi/10.1103/PhysRevLett.79.2113}.

\bibitem{barisic2013i}
\bibinfo{author}{Bari{\v{s}}i{\'c}, N.}, \bibinfo{author}{Chan, M.~K.},
  \bibinfo{author}{Li, Y.}, \bibinfo{author}{Greven, M.} \emph{et~al.}
\newblock \bibinfo{title}{Universal sheet resistance and revised phase diagram
  of the cuprate high-temperature superconductors}.
\newblock \emph{\bibinfo{journal}{Proceedings of the National Academy of
  Sciences}} \textbf{\bibinfo{volume}{110}}, \bibinfo{pages}{12235--12240}
  (\bibinfo{year}{2013}).
\newblock \bibinfo{note}{Edited by J. C. Seamus Davis, Cornell University,
  Ithaca, NY}.

\bibitem{giraldogallo2018}
\bibinfo{author}{Giraldo-Gallo, P.} \emph{et~al.}
\newblock \bibinfo{title}{Scale-invariant magnetoresistance in a cuprate
  superconductor}.
\newblock \emph{\bibinfo{journal}{Science}} \textbf{\bibinfo{volume}{361}},
  \bibinfo{pages}{479--481} (\bibinfo{year}{2018}).

\bibitem{cooper2009}
\bibinfo{author}{Cooper, R.~A.} \emph{et~al.}
\newblock \bibinfo{title}{Anomalous criticality in the electrical resistivity
  of lsco}.
\newblock \emph{\bibinfo{journal}{Science}} \textbf{\bibinfo{volume}{323}},
  \bibinfo{pages}{603} (\bibinfo{year}{2009}).

\bibitem{proust2019}
\bibinfo{author}{Proust, C.} \& \bibinfo{author}{Taillefer, L.}
\newblock \bibinfo{title}{The remarkable underlying ground states of cuprate
  superconductors}.
\newblock \emph{\bibinfo{journal}{Annual Review of Condensed Matter Physics}}
  \textbf{\bibinfo{volume}{10}}, \bibinfo{pages}{409--429}
  (\bibinfo{year}{2019}).
\newblock
  \urlprefix\url{https://doi.org/10.1146/annurev-conmatphys-031218-013210}.

\bibitem{ashcroft1976}
\bibinfo{author}{Ashcroft, N.~W.} \& \bibinfo{author}{Mermin, N.~D.}
\newblock \emph{\bibinfo{title}{Solid State Physics}}
  (\bibinfo{publisher}{Saunders College Publishing},
  \bibinfo{address}{Orlando}, \bibinfo{year}{1976}).

\bibitem{loram2001}
\bibinfo{author}{Loram, J.~W.} \emph{et~al.}
\newblock \bibinfo{title}{Evidence on the pseudogap and condensate from the
  electronic specific heat}.
\newblock \emph{\bibinfo{journal}{J. Physics and Chemistry of Solids}}
  \textbf{\bibinfo{volume}{62}}, \bibinfo{pages}{59--64}
  (\bibinfo{year}{2001}).

\bibitem{tallon2022}
\bibinfo{author}{Tallon, J.~L.} \& \bibinfo{author}{Storey, J.~G.}
\newblock \bibinfo{title}{Thermodynamics of the pseudogap in cuprates}.
\newblock \emph{\bibinfo{journal}{Frontiers in Physics}}
  \textbf{\bibinfo{volume}{10}}, \bibinfo{pages}{1030616}
  (\bibinfo{year}{2022}).

\bibitem{ando2004a}
\bibinfo{author}{Ando, Y.}, \bibinfo{author}{Kurita, Y.},
  \bibinfo{author}{Komiya, S.}, \bibinfo{author}{Ono, S.} \&
  \bibinfo{author}{Segawa, K.}
\newblock \bibinfo{title}{Evolution of the hall coefficient and the peculiar
  electronic structure of the cuprate superconductors}.
\newblock \emph{\bibinfo{journal}{Phys. Rev. Lett.}}
  \textbf{\bibinfo{volume}{92}}, \bibinfo{pages}{197001}
  (\bibinfo{year}{2004}).
\newblock
  \urlprefix\url{https://link.aps.org/doi/10.1103/PhysRevLett.92.197001}.

\bibitem{ono2007}
\bibinfo{author}{Ono, S.}, \bibinfo{author}{Komiya, S.} \&
  \bibinfo{author}{Ando, Y.}
\newblock \bibinfo{title}{Strong charge fluctuations manifested in the
  high-temperature hall coefficient of high-${T}_{c}$ cuprates}.
\newblock \emph{\bibinfo{journal}{Phys. Rev. B}} \textbf{\bibinfo{volume}{75}},
  \bibinfo{pages}{024515} (\bibinfo{year}{2007}).
\newblock \urlprefix\url{https://link.aps.org/doi/10.1103/PhysRevB.75.024515}.

\bibitem{horio2018}
\bibinfo{author}{Horio, M.} \emph{et~al.}
\newblock \bibinfo{title}{Three-dimensional fermi surface of overdoped la-based
  cuprates}.
\newblock \emph{\bibinfo{journal}{Phys. Rev. Lett.}}
  \textbf{\bibinfo{volume}{121}}, \bibinfo{pages}{077004}
  (\bibinfo{year}{2018}).

\bibitem{timusk1999}
\bibinfo{author}{Timusk, T.} \& \bibinfo{author}{Statt, B.}
\newblock \bibinfo{title}{The pseudogap in high-temperature superconductors: an
  experimental survey}.
\newblock \emph{\bibinfo{journal}{Rep. Prog. Phys.}}
  \textbf{\bibinfo{volume}{62}}, \bibinfo{pages}{61--122}
  (\bibinfo{year}{1999}).

\bibitem{grissonnanche2021}
\bibinfo{author}{Grissonnanche, G.} \emph{et~al.}
\newblock \bibinfo{title}{Linear-in temperature resistivity from an isotropic
  planckian scattering rate}.
\newblock \emph{\bibinfo{journal}{Nature}} \textbf{\bibinfo{volume}{595}},
  \bibinfo{pages}{667--672} (\bibinfo{year}{2021}).

\bibitem{ataei2022}
\bibinfo{author}{Ataei, A.} \emph{et~al.}
\newblock \bibinfo{title}{Electrons with planckian scattering obey standard
  orbital motion in a magnetic field}.
\newblock \emph{\bibinfo{journal}{Nature Physics}}
  \textbf{\bibinfo{volume}{18}}, \bibinfo{pages}{1420--1424}
  (\bibinfo{year}{2022}).

\bibitem{yoshida2006}
\bibinfo{author}{Yoshida, T.} \emph{et~al.}
\newblock \bibinfo{title}{Systematic doping evolution of the underlying fermi
  surface of la$_{2-x}$sr$_x$cuo$_4$}.
\newblock \emph{\bibinfo{journal}{Physical Review B}}
  \textbf{\bibinfo{volume}{74}}, \bibinfo{pages}{224510}
  (\bibinfo{year}{2006}).

\bibitem{momono2002}
\bibinfo{author}{Momono, N.}, \bibinfo{author}{Matsuzaki, T.},
  \bibinfo{author}{Oda, M.} \& \bibinfo{author}{Ido, M.}
\newblock \bibinfo{title}{Superconducting condensation energy and pseudogap
  formation in lsco: New energy scale for superconductivity}.
\newblock \emph{\bibinfo{journal}{Journal of the Physical Society of Japan}}
  \textbf{\bibinfo{volume}{71}}, \bibinfo{pages}{2832--2835}
  (\bibinfo{year}{2002}).

\bibitem{badoux2016}
\bibinfo{author}{Badoux, S.} \emph{et~al.}
\newblock \bibinfo{title}{Change of carrier density at the pseudogap critical
  point of a cuprate superconductor}.
\newblock \emph{\bibinfo{journal}{Nature}} \textbf{\bibinfo{volume}{531}},
  \bibinfo{pages}{210} (\bibinfo{year}{2016}).

\bibitem{laliberte2016}
\bibinfo{author}{Lalibert{\'e}, F.} \emph{et~al.}
\newblock \bibinfo{title}{Origin of the metal-to-insulator crossover in cuprate
  superconductors}.
\newblock \emph{\bibinfo{journal}{arXiv preprint arXiv:1606.04491}}
  (\bibinfo{year}{2016}).
\newblock \eprint{1606.04491}.

\bibitem{michon2019}
\bibinfo{author}{Michon, B.} \emph{et~al.}
\newblock \bibinfo{title}{Thermodynamic signatures of quantum criticality in
  cuprate superconductors}.
\newblock \emph{\bibinfo{journal}{Nature}} \textbf{\bibinfo{volume}{567}},
  \bibinfo{pages}{218--222} (\bibinfo{year}{2019}).

\bibitem{putzke2021}
\bibinfo{author}{Putzke, C.} \emph{et~al.}
\newblock \bibinfo{title}{Reduced hall carrier density in the overdoped strange
  metal regime of cuprate superconductors}.
\newblock \emph{\bibinfo{journal}{Nature Physics}}
  \textbf{\bibinfo{volume}{17}}, \bibinfo{pages}{826--831}
  (\bibinfo{year}{2021}).

\bibitem{girod2021}
\bibinfo{author}{Girod, C.} \emph{et~al.}
\newblock \bibinfo{title}{Normal state specific heat in the cuprate
  superconductors lsco near the critical point of the pseudogap phase}.
\newblock \emph{\bibinfo{journal}{Phys. Rev. B}}
  \textbf{\bibinfo{volume}{103}}, \bibinfo{pages}{214506}
  (\bibinfo{year}{2021}).

\bibitem{shekhter2022}
\bibinfo{author}{Shekhter, A.} \emph{et~al.}
\newblock \bibinfo{title}{Energy-scale competition in the hall resistivity of a
  strange metal}.
\newblock \emph{\bibinfo{journal}{arXiv preprint arXiv:2207.10244}}
  (\bibinfo{year}{2022}).

\bibitem{legros2022}
\bibinfo{author}{Legros, A.} \emph{et~al.}
\newblock \bibinfo{title}{Evolution of the cyclotron mass with doping in lsco}.
\newblock \emph{\bibinfo{journal}{Physical Review B}}
  \textbf{\bibinfo{volume}{106}}, \bibinfo{pages}{195110}
  (\bibinfo{year}{2022}).
\newblock \bibinfo{note}{Editors' Suggestion}.

\bibitem{padilla2005}
\bibinfo{author}{Padilla, W.~J.} \emph{et~al.}
\newblock \bibinfo{title}{Constant effective mass across the phase diagram of
  high-${T}_{c}$ cuprates}.
\newblock \emph{\bibinfo{journal}{Phys. Rev. B}} \textbf{\bibinfo{volume}{72}},
  \bibinfo{pages}{060511} (\bibinfo{year}{2005}).
\newblock \urlprefix\url{https://link.aps.org/doi/10.1103/PhysRevB.72.060511}.

\bibitem{michon2021}
\bibinfo{author}{Michon, B.} \emph{et~al.}
\newblock \bibinfo{title}{Spectral weight of hole-doped cuprates across the
  pseudogap critical point}.
\newblock \emph{\bibinfo{journal}{Physical Review Research}}
  \textbf{\bibinfo{volume}{3}}, \bibinfo{pages}{043125} (\bibinfo{year}{2021}).

\bibitem{vollhardt1984}
\bibinfo{author}{Vollhardt, D.}
\newblock \bibinfo{title}{Normal $^3$he: an almost localized fermi liquid}.
\newblock \emph{\bibinfo{journal}{Rev. Mod. Phys.}}
  \textbf{\bibinfo{volume}{56}}, \bibinfo{pages}{99} (\bibinfo{year}{1984}).

\bibitem{vollhardt1987}
\bibinfo{author}{Vollhardt, D.}, \bibinfo{author}{Wolfle, P.} \&
  \bibinfo{author}{Anderson, P.~W.}
\newblock \bibinfo{title}{Gutzwiller-hubbard lattice-gas model with variable
  density: Application to normal liquid $^3$he}.
\newblock \emph{\bibinfo{journal}{Phys. Rev. B}} \textbf{\bibinfo{volume}{35}},
  \bibinfo{pages}{6703--6715} (\bibinfo{year}{1987}).

\bibitem{anderson1987}
\bibinfo{author}{Anderson, P.~W.}
\newblock \bibinfo{title}{The resonating valence bond state in lsco and
  superconductivity}.
\newblock \emph{\bibinfo{journal}{Science}} \textbf{\bibinfo{volume}{235}},
  \bibinfo{pages}{1196--1198} (\bibinfo{year}{1987}).

\bibitem{mackenzie1996}
\bibinfo{author}{Mackenzie, A.~P.}, \bibinfo{author}{Julian, S.~R.},
  \bibinfo{author}{Sinclair, D.~C.} \& \bibinfo{author}{Lin, C.~T.}
\newblock \bibinfo{title}{Normal-state magnetotransport in superconducting
  tl2201 to millikelvin temperatures}.
\newblock \emph{\bibinfo{journal}{Phys. Rev. B}} \textbf{\bibinfo{volume}{53}},
  \bibinfo{pages}{5848} (\bibinfo{year}{1996}).
\newblock \urlprefix\url{https://doi.org/10.1103/PhysRevB.53.5848}.

\bibitem{hussey2011}
\bibinfo{author}{Hussey, N.} \emph{et~al.}
\newblock \bibinfo{title}{Dichotomy in the t-linear resistivity in hole-doped
  cuprates}.
\newblock \emph{\bibinfo{journal}{Philosophical Transactions of the Royal
  Society A: Mathematical, Physical and Engineering Sciences}}
  \textbf{\bibinfo{volume}{369}}, \bibinfo{pages}{1626--1639}
  (\bibinfo{year}{2011}).

\bibitem{finkelstein1990}
\bibinfo{author}{Finkel'stein, A.~M.}
\newblock \bibinfo{title}{Electron liquid in disordered conductors}.
\newblock In \bibinfo{editor}{Khalatnikov, I.~M.} (ed.)
  \emph{\bibinfo{booktitle}{Soviet Scientific Reviews, Section A: Physics
  Reviews}}, vol. \bibinfo{volume}{14, Part 2}, \bibinfo{pages}{1--101}
  (\bibinfo{publisher}{CRC Press}, \bibinfo{year}{1990}).

\bibitem{finkelstein1983}
\bibinfo{author}{Finkel'shte{\i}n, A.~M.}
\newblock \bibinfo{title}{Influence of coulomb interaction on the properties of
  disordered metals}.
\newblock \emph{\bibinfo{journal}{Zh. Eksp. Teor. Fiz.}}
  \textbf{\bibinfo{volume}{84}}, \bibinfo{pages}{168--189}
  (\bibinfo{year}{1983}).
\newblock \bibinfo{note}{Submitted 17 June 1982}.

\bibitem{stricker2014}
\bibinfo{author}{Stricker, D.} \emph{et~al.}
\newblock \bibinfo{title}{Optical response of sr$_2$ruo$_4$ reveals universal
  fermi-liquid scaling and quasiparticles beyond landau theory}.
\newblock \emph{\bibinfo{journal}{Physical Review Letters}}
  \textbf{\bibinfo{volume}{113}}, \bibinfo{pages}{087404}
  (\bibinfo{year}{2014}).

\bibitem{zaanen2020}
\bibinfo{author}{Zaanen, J.}
\newblock \bibinfo{title}{Carriers that count}.
\newblock \emph{\bibinfo{journal}{Nature Physics}}
  \textbf{\bibinfo{volume}{16}}, \bibinfo{pages}{1171--1172}
  (\bibinfo{year}{2020}).

\bibitem{basov2005}
\bibinfo{author}{Basov, D.~N.} \& \bibinfo{author}{Timusk, T.}
\newblock \bibinfo{title}{Electrodynamics of high-\( t_c \) superconductors}.
\newblock \emph{\bibinfo{journal}{Reviews of Modern Physics}}
  \textbf{\bibinfo{volume}{77}}, \bibinfo{pages}{721--779}
  (\bibinfo{year}{2005}).

\bibitem{armitage2009}
\bibinfo{author}{Armitage, N.~P.}
\newblock \bibinfo{title}{Electrodynamics of correlated electron systems}
  (\bibinfo{year}{2009}).
\newblock \urlprefix\url{https://arxiv.org/abs/0908.1126}.
\newblock \bibinfo{note}{ArXiv:0908.1126 [cond-mat.str-el]},
  \eprint{0908.1126}.

\bibitem{liang2006}
\bibinfo{author}{Liang, R.}, \bibinfo{author}{Bonn, D.~A.} \&
  \bibinfo{author}{Hardy, W.~N.}
\newblock \bibinfo{title}{Evaluation of ${\mathrm{cuo}}_{2}$ plane hole doping
  in ${\mathrm{yba}}_{2}{\mathrm{cu}}_{3}{\mathrm{o}}_{6+x}$ single crystals}.
\newblock \emph{\bibinfo{journal}{Phys. Rev. B}} \textbf{\bibinfo{volume}{73}},
  \bibinfo{pages}{180505} (\bibinfo{year}{2006}).
\newblock \urlprefix\url{https://link.aps.org/doi/10.1103/PhysRevB.73.180505}.

\bibitem{vanheumen2022}
\bibinfo{author}{van Heumen, E.} \emph{et~al.}
\newblock \bibinfo{title}{Strange metal electrodynamics across the phase
  diagram of biscco cuprates}.
\newblock \emph{\bibinfo{journal}{Physical Review B}}
  \textbf{\bibinfo{volume}{106}}, \bibinfo{pages}{054515}
  (\bibinfo{year}{2022}).

\bibitem{uemura1989}
\bibinfo{author}{Uemura, Y.~J.} \emph{et~al.}
\newblock \bibinfo{title}{Universal correlations between ${T}_{c}$ and
  $\frac{{n}_{s}}{{m}^{*}}$ (carrier density over effective mass) in
  high-${T}_{c}$ cuprate superconductors}.
\newblock \emph{\bibinfo{journal}{Phys. Rev. Lett.}}
  \textbf{\bibinfo{volume}{62}}, \bibinfo{pages}{2317--2320}
  (\bibinfo{year}{1989}).
\newblock \urlprefix\url{https://link.aps.org/doi/10.1103/PhysRevLett.62.2317}.

\bibitem{uemura1991}
\bibinfo{author}{Uemura, Y.~J.} \emph{et~al.}
\newblock \bibinfo{title}{Basic similarities among cuprate, bismuthate,
  organic, chevrel-phase, and heavy-fermion superconductors shown by
  penetration-depth measurements}.
\newblock \emph{\bibinfo{journal}{Physical Review Letters}}
  \textbf{\bibinfo{volume}{66}}, \bibinfo{pages}{2665--2668}
  (\bibinfo{year}{1991}).

\bibitem{rourke2010}
\bibinfo{author}{Rourke, P. M.~C.} \emph{et~al.}
\newblock \bibinfo{title}{A detailed de haas van alphen effect study of the
  overdoped cuprate tl2201}.
\newblock \emph{\bibinfo{journal}{New J. Phys.}} \textbf{\bibinfo{volume}{12}},
  \bibinfo{pages}{105009} (\bibinfo{year}{2010}).

\bibitem{brinkman1970}
\bibinfo{author}{Brinkman, W.~F.} \& \bibinfo{author}{Rice, T.~M.}
\newblock \bibinfo{title}{Application of gutzwiller's variational method to the
  metal-insulator transition}.
\newblock \emph{\bibinfo{journal}{Phys. Rev. B}} \textbf{\bibinfo{volume}{2}},
  \bibinfo{pages}{4302--4304} (\bibinfo{year}{1970}).

\bibitem{phillips2020}
\bibinfo{author}{Phillips, P.~W.}, \bibinfo{author}{Yeo, L.} \&
  \bibinfo{author}{Huang, E.~W.}
\newblock \bibinfo{title}{Exact theory for superconductivity in a doped mott
  insulator}.
\newblock \emph{\bibinfo{journal}{Nature Physics}}
  \textbf{\bibinfo{volume}{16}}, \bibinfo{pages}{1175--1180}
  (\bibinfo{year}{2020}).

\bibitem{leggett1968}
\bibinfo{author}{Leggett, A.~J.}
\newblock \bibinfo{title}{Inequalities, instabilities, and renormalization in
  metals and other fermi liquids}.
\newblock \emph{\bibinfo{journal}{Annals of Physics}}
  \textbf{\bibinfo{volume}{46}}, \bibinfo{pages}{76--113}
  (\bibinfo{year}{1968}).

\bibitem{leggett1975}
\bibinfo{author}{Leggett, A.~J.}
\newblock \bibinfo{title}{A theoretical description of the new phases of liquid
  $^3$he}.
\newblock \emph{\bibinfo{journal}{Rev. Mod. Phys.}}
  \textbf{\bibinfo{volume}{47}}, \bibinfo{pages}{331--414}
  (\bibinfo{year}{1975}).

\bibitem{wheatley1975}
\bibinfo{author}{Wheatley, J.~C.}
\newblock \bibinfo{title}{Experimental properties of superfluid $^3$he}.
\newblock \emph{\bibinfo{journal}{Rev. Mod. Phys.}}
  \textbf{\bibinfo{volume}{47}}, \bibinfo{pages}{415--470}
  (\bibinfo{year}{1975}).

\bibitem{doniach1966}
\bibinfo{author}{Doniach, S.} \& \bibinfo{author}{Engelsberg, S.}
\newblock \emph{\bibinfo{journal}{Phys. Rev. Lett.}}
  \textbf{\bibinfo{volume}{17}}, \bibinfo{pages}{750--753}
  (\bibinfo{year}{1966}).

\bibitem{armitage1975helium}
\bibinfo{author}{Anderson, P.~W.} \& \bibinfo{author}{Brinkman, W.~F.}
\newblock \emph{\bibinfo{title}{The Helium Liquids: Proceedings of the
  Fifteenth Scottish Universities Summer School in Physics, 1974}}.
\newblock NATO Advanced Study Institutes (\bibinfo{publisher}{Academic Press},
  \bibinfo{year}{1975}).

\bibitem{andreev1969}
\bibinfo{author}{Andreev, A.~F.} \& \bibinfo{author}{Lifshitz, I.~M.}
\newblock \bibinfo{title}{Quantum theory of defects in crystals}.
\newblock \emph{\bibinfo{journal}{Soviet Physics JETP}}
  \textbf{\bibinfo{volume}{29}}, \bibinfo{pages}{1107--1113}
  (\bibinfo{year}{1969}).
\newblock \bibinfo{note}{Zh. Eksp. Teor. Fiz. 56, 2057-2068 (June, 1969)}.

\bibitem{grilly1971}
\bibinfo{author}{Grilly, E.~R.}
\newblock \bibinfo{title}{Pressure-volume-temperature relations in liquid and
  solid $^3$he}.
\newblock \emph{\bibinfo{journal}{Journal of Low Temperature Physics}}
  \textbf{\bibinfo{volume}{4}}, \bibinfo{pages}{615--635}
  (\bibinfo{year}{1971}).
\newblock \urlprefix\url{https://doi.org/10.1007/BF00628297}.

\bibitem{landau1980}
\bibinfo{author}{Landau, L.~D.} \& \bibinfo{author}{Lifshitz, E.~M.}
\newblock \emph{\bibinfo{title}{Statistical Physics, Third Edition, Part 1:
  Volume 5 (Course of Theoretical Physics, Volume 5)}},
  vol.~\bibinfo{volume}{5} of \emph{\bibinfo{series}{Course of Theoretical
  Physics}} (\bibinfo{publisher}{Butterworth-Heinemann},
  \bibinfo{address}{Oxford}, \bibinfo{year}{1980}), \bibinfo{edition}{3rd} edn.

\bibitem{mattheiss1987}
\bibinfo{author}{Mattheiss, L.~F.}
\newblock \bibinfo{title}{Electronic band properties and superconductivity in
  lsco}.
\newblock \emph{\bibinfo{journal}{Phys. Rev. Lett.}}
  \textbf{\bibinfo{volume}{58}}, \bibinfo{pages}{1028} (\bibinfo{year}{1987}).

\bibitem{singh1994}
\bibinfo{author}{Singh, D.~J.} \& \bibinfo{author}{Pickett, W.~E.}
\newblock \bibinfo{title}{Electronic structure studies of doped and undoped
  hg-ba-ca-cu-o}.
\newblock \emph{\bibinfo{journal}{Physica C}}
  \textbf{\bibinfo{volume}{C235-240}}, \bibinfo{pages}{2113--2114}
  (\bibinfo{year}{1994}).

\bibitem{pavarini2001}
\bibinfo{author}{Pavarini, E.}, \bibinfo{author}{Dasgupta, I.},
  \bibinfo{author}{Saha-Dasgupta, T.}, \bibinfo{author}{Jepsen, O.} \&
  \bibinfo{author}{Andersen, O.~K.}
\newblock \bibinfo{title}{Band-structure trend in hole-doped cuprates and
  correlation with $t_{\rm c,max}$}.
\newblock \emph{\bibinfo{journal}{Phys. Rev. Lett.}}
  \textbf{\bibinfo{volume}{87}}, \bibinfo{pages}{047003}
  (\bibinfo{year}{2001}).

\bibitem{zou1988}
\bibinfo{author}{Zou, Z.} \& \bibinfo{author}{Anderson, P.~W.}
\newblock \bibinfo{title}{Neutral fermion, charge-e boson excitations in the
  resonating-valence-bond state and superconductivity in la2cuo4-based
  compounds}.
\newblock \emph{\bibinfo{journal}{Physical Review B}}
  \textbf{\bibinfo{volume}{37}}, \bibinfo{pages}{627--630}
  (\bibinfo{year}{1988}).

\bibitem{basov2011}
\bibinfo{author}{Basov, D.~N.}, \bibinfo{author}{Averitt, R.~D.},
  \bibinfo{author}{van~der Marel, D.}, \bibinfo{author}{Dressel, M.} \&
  \bibinfo{author}{Haule, K.}
\newblock \bibinfo{title}{Electrodynamics of correlated electron materials}.
\newblock \emph{\bibinfo{journal}{Rev. Mod. Phys.}}
  \textbf{\bibinfo{volume}{83}}, \bibinfo{pages}{471--541}
  (\bibinfo{year}{2011}).
\newblock \urlprefix\url{https://link.aps.org/doi/10.1103/RevModPhys.83.471}.

\bibitem{luttinger1960}
\bibinfo{author}{Luttinger, J.~M.} \& \bibinfo{author}{Ward, J.~C.}
\newblock \bibinfo{title}{Ground-state energy of a many-fermion system. ii}.
\newblock \emph{\bibinfo{journal}{Physical Review}}
  \textbf{\bibinfo{volume}{118}}, \bibinfo{pages}{1417--1427}
  (\bibinfo{year}{1960}).

\bibitem{ramshaw2015}
\bibinfo{author}{Ramshaw, B.~J.} \emph{et~al.}
\newblock \bibinfo{title}{Quasiparticle mass enhancement approaching optimal
  doping in a high-$t_c$ superconductor}.
\newblock \emph{\bibinfo{journal}{Science}} \textbf{\bibinfo{volume}{348}},
  \bibinfo{pages}{317--320} (\bibinfo{year}{2015}).

\bibitem{varma1999}
\bibinfo{author}{Varma, C.~M.}
\newblock \bibinfo{title}{Pseudogap phase and the quantum-critical point in
  copper-oxide metals}.
\newblock \emph{\bibinfo{journal}{Physical Review Letters}}
  \textbf{\bibinfo{volume}{83}}, \bibinfo{pages}{3538--3541}
  (\bibinfo{year}{1999}).

\bibitem{ayres2021}
\bibinfo{author}{Ayres, J.} \emph{et~al.}
\newblock \bibinfo{title}{Incoherent transport across the strange-metal regime
  of overdoped cuprates}.
\newblock \emph{\bibinfo{journal}{Nature}} \textbf{\bibinfo{volume}{595}},
  \bibinfo{pages}{661--666} (\bibinfo{year}{2021}).

\bibitem{ramshaw2026}
\bibinfo{author}{Ramshaw, B.~J.} \& \bibinfo{author}{Kivelson, S.~A.}
\newblock \bibinfo{title}{Superconductivity in overdoped cuprates can be
  understood from a bcs perspective!}
\newblock \emph{\bibinfo{journal}{Nature Reviews Physics}}
  (\bibinfo{year}{2026}).
\newblock \bibinfo{note}{In press; arXiv:2510.25767}, \eprint{2510.25767}.

\bibitem{rullieralbenque2008}
\bibinfo{author}{Rullier-Albenque, F.}, \bibinfo{author}{Alloul, H.},
  \bibinfo{author}{Balakirev, F.} \& \bibinfo{author}{Proust, C.}
\newblock \bibinfo{title}{Disorder, metal-insulator crossover and phase diagram
  in high-tc cuprates}.
\newblock \emph{\bibinfo{journal}{Europhysics Letters}}
  \textbf{\bibinfo{volume}{81}}, \bibinfo{pages}{37008} (\bibinfo{year}{2008}).
\newblock \urlprefix\url{https://dx.doi.org/10.1209/0295-5075/81/37008}.

\bibitem{kotliar1988}
\bibinfo{author}{Kotliar, G.} \& \bibinfo{author}{Liu, J.}
\newblock \bibinfo{title}{Superexchange mechanism and $d$-wave
  superconductivity}.
\newblock \emph{\bibinfo{journal}{Phys. Rev. B}} \textbf{\bibinfo{volume}{38}},
  \bibinfo{pages}{5142--5145} (\bibinfo{year}{1988}).

\bibitem{luttinger1960a}
\bibinfo{author}{Luttinger, J.~M.} \& \bibinfo{author}{Ward, J.~C.}
\newblock \bibinfo{title}{Ground-state energy of a many-fermion system. ii}.
\newblock \emph{\bibinfo{journal}{Phys. Rev.}} \textbf{\bibinfo{volume}{118}},
  \bibinfo{pages}{1417--1427} (\bibinfo{year}{1960}).

\bibitem{luttinger1960b}
\bibinfo{author}{Luttinger, J.~M.}
\newblock \bibinfo{title}{Fermi surface and some simple equilibrium properties
  of a system of interacting fermions}.
\newblock \emph{\bibinfo{journal}{Phys. Rev.}} \textbf{\bibinfo{volume}{119}},
  \bibinfo{pages}{1153--1163} (\bibinfo{year}{1960}).

\bibitem{zhong2022}
\bibinfo{author}{Zhong, Y.} \emph{et~al.}
\newblock \bibinfo{title}{Differentiated roles of lifshitz transition on
  thermodynamics and superconductivity in lsco}.
\newblock \emph{\bibinfo{journal}{Proc. Natl. Acad. USA}}
  \textbf{\bibinfo{volume}{119}}, \bibinfo{pages}{e2204630119}
  (\bibinfo{year}{2022}).

\bibitem{musser2022}
\bibinfo{author}{Musser, S.}, \bibinfo{author}{Chowdhury, D.},
  \bibinfo{author}{Lee, P.~A.} \& \bibinfo{author}{Senthil, T.}
\newblock \bibinfo{title}{Interpreting angle-dependent magnetoresistance in
  layered materials: Application to cuprates}.
\newblock \emph{\bibinfo{journal}{Phys. Rev. B}}
  \textbf{\bibinfo{volume}{105}}, \bibinfo{pages}{125105}
  (\bibinfo{year}{2022}).
\newblock \urlprefix\url{https://link.aps.org/doi/10.1103/PhysRevB.105.125105}.

\bibitem{vandermarel2003}
\bibinfo{author}{van~der Marel, D.} \emph{et~al.}
\newblock \bibinfo{title}{Quantum critical behaviour in a high-{$T_c$}
  superconductor}.
\newblock \emph{\bibinfo{journal}{Nature}} \textbf{\bibinfo{volume}{425}},
  \bibinfo{pages}{271--274} (\bibinfo{year}{2003}).

\end{thebibliography}
%

\end{document}